\documentclass[twocolumn, prl, aps, superscriptaddress,showpacs,amsmath,amssymb,floatfix]{revtex4-1}                                                                                              
\usepackage{hyperref}
\usepackage{amssymb}
\usepackage{amsmath}
\usepackage{epsfig}
\usepackage{color}
\usepackage{tabu}
\usepackage{mathtools}
\usepackage{physics}
\usepackage{float}
\usepackage{diagbox}
\usepackage{amssymb}
\usepackage{amssymb}
\usepackage{graphicx}
\usepackage{amsmath}
\usepackage{amsfonts}
\usepackage{array}
\usepackage{multirow}
\usepackage{color}
\usepackage{esint}
\usepackage{bm}
\usepackage{color}
\usepackage{adjustbox}
\usepackage{titlesec}
\def\r{\mathbf{r}}
\usepackage[normalem]{ulem}
\newcommand{\Roma}[1]
{\MakeUppercase{\romannumeral #1}}
\begin{document}
	
	\title{Interplay between non-Hermiticity and non-Abelian gauge potential in topological photonics}
	\author{Jia-Qi Cai}
	\affiliation{School of Physics, Huazhong University of Science and Technology, Wuhan 430074, China}
	\author{Qing-Yun Yang}
    \affiliation{School of Physics, Huazhong University of Science and Technology, Wuhan 430074, China}
	\author{Zheng-Yuan Xue}
	\affiliation{Guangdong Provincial Key Laboratory of Quantum Engineering and Quantum Materials, and School of Physics and Telecommunication Engineering, South China Normal University, Guangzhou 510006, China}
	\author{Ming Gong}
	\email{gongm@ustc.edu.cn}
	\affiliation{CAS Key Laboratory of Quantum Information, University of Science and Technology of China, Hefei 230026, China}
	\affiliation{CAS Center For Excellence in Quantum Information and Quantum Physics}
	\affiliation{Synergetic Innovation Center of Quantum Information and Quantum Physics, University of Science and Technology of China, Hefei, 230026, P.R. China}
	\author{Guang-Can Guo}
	\affiliation{CAS Key Laboratory of Quantum Information, University of Science and Technology of China, Hefei 230026, China}
	\affiliation{CAS Center For Excellence in Quantum Information and Quantum Physics}
	\affiliation{Synergetic Innovation Center of Quantum Information and Quantum Physics, University of Science and Technology of China, Hefei, 230026, P.R. China}
	\author{Yong Hu} \email{huyong@hust.edu.cn}
	\affiliation{School of Physics, Huazhong University of Science and Technology, Wuhan 430074, China}
	\date{\today}
	
	\begin{abstract}
		Topological phases in spinless non-Hermitian models have been widely studied both theoretically and experimentally in some artificial materials
		using photonics, phononics and magnon. In this work, we investigate the interplay between non-Hermitian loss and gain and 
		non-Abelian gauge potential realized in a two-component superconducting circuit. In our model, the non-Hermiticity 
		along only gives rise to trivial gain and loss to the states; while the non-Abelian gauge along gives rise to flying butterfly spectra 
		and associated edge modes, which in photonics can be directly measured by the intensity of photons at the boundaries. These two terms do not commute, and their interplay can 
		give rise to several intriguing non-Hermitian phases, including the fully gapped quantum spin Hall (QSH) 
		phase, gapless QSH phase, trivial gapped phase and gapless metallic phase. The bulk-edge correspondence is absent and we find that during the closing of 
		energy gap in the gapped QSH phase, the system enters the gapless QSH phase regime which still supports two counter-propagating edge modes. We have also unveiled
		the intriguing role of non-Hermiticity on the chiral symmetry and time-reversal symmetry of the Hermitian models, which can be applied to other physical models. 
	\end{abstract}
	
	\maketitle

	The study of topological matters in physics initiated from quantum Hall effect~\cite{klitzing1980new, tsui1982two, laughlin1983} is of fundamental importance in modern physics.
	A lot of topological insulators and superconductors have been realized using solid materials~\cite{Ando2013topological, ando2015topological,hasan2010colloquium}, which 
	can lead to interesting applications such as spintronics and topological quantum computation\cite{murakami2003dissipationless, koralek2009emergence, yuan2014generation, 
	hsieh2009observation, sau2010generic, hasan2010colloquium}. Moreover, the state-of-art technologies still allow us to simulate these novel phases 
	in the context of quantum simulation in a more controllable manner~\cite{Georgescu2014}. Recently, this idea has been implemented in various systems, including photonics~\cite{sala2015spin,lu2014topological,ozawa2018topological,khanikaev2013photonic,cheng2016robust}, magnons~\cite{zhang2013topological,mochizuki2014thermally}, 
	ultracold atoms~\cite{aidelsburger2013realization, Goldman2009, aidelsburger2011experimental,aidelsburger2015measuring,Ningyuan2015}, linear circuits ~\cite{imhof2018topolectrical, Zhu2018}, and superconducting circuits~\cite{Roushan2017a}, {\it etc}. Due to the salient advantages in experiments, the details of these topological protected edge modes, such as the zero modes, Fermi points 
	and Fermi arcs, can be seen much clearer as compared with those in solid materials~\cite{lu2015experimental,yang2018ideal}. 
	These achievements have greatly advanced our understanding of topological matters and may seek for new applications ~\cite{st2017lasing, harari2018topological}.
	
	Here we are mainly interested in these topological phases in the superconducting circuit, due to its long coherence time~\cite{Paik2011} and its scalability in industrial fabrication. 
	To date, superconducting circuits up to 92 physical qubits have been reported to achieve the quantum supremacy~\cite{harrow2017quantum, song201710}. The topological phases with these 
	circuits can be realized with much larger size due to the much lower threshold of uniformity. This platform is also intriguing for its controllability in loss and gain~\cite{Quijandria2018, 
	feng2017non, regensburger2012parity,schomerus2013topologically} for the realization of non-Hermitian topological phases. 

	\begin{figure}
		\centering
		\includegraphics[width=0.49\textwidth]{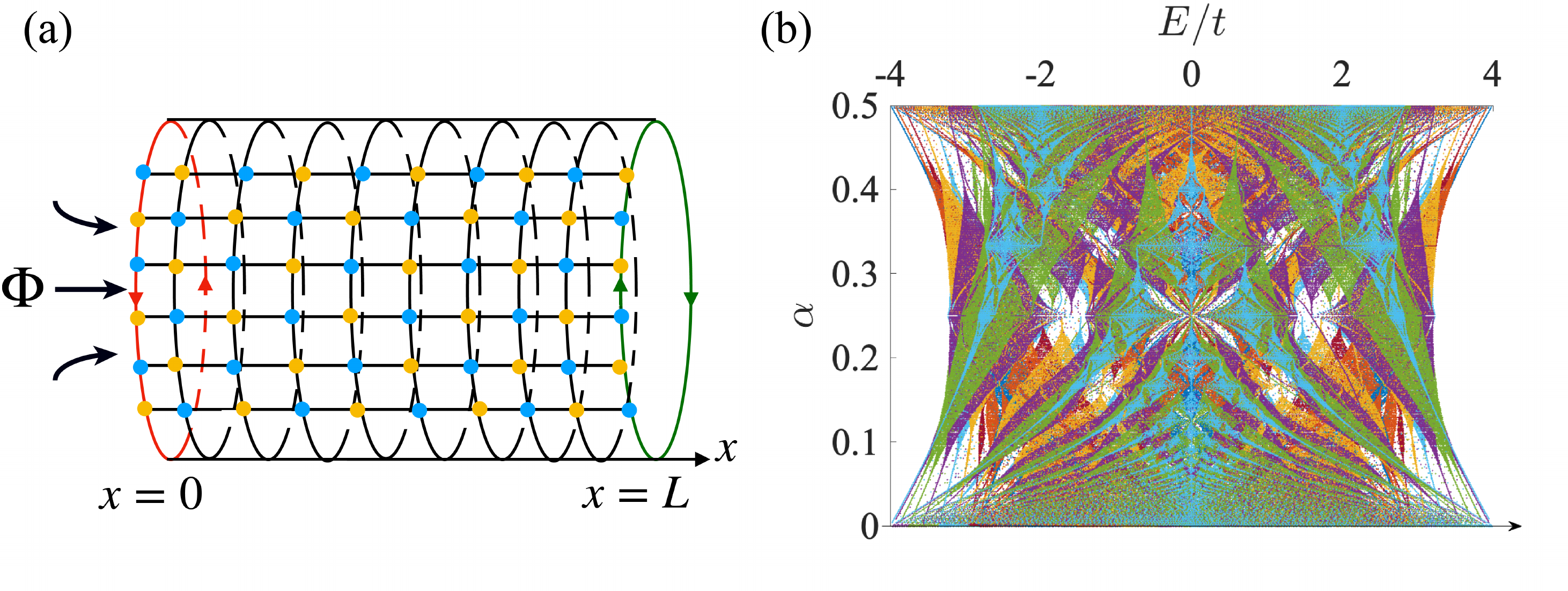}
		\caption{(a) Periodic boundary condition for the realization of Hofstadter butterfly effect. 
		(b) The flying Hofstadter spectra. In this plot the momentum $k_y$ along the cylindrical geometry (a) 
		plays the role of threaded flux $\Phi$. In this plot $\gamma = 1/4$ to achieve maximal coupling between the two components.}
		\label{fig-fig1}
	\end{figure}

	In previous literature, the topological non-Hermitian models were generally considered with some spinless two-band models~\cite{Shen2018,Yao2018a,Yao2018,schomerus2013parity}. Here we are interested in the basic question that how non-Abelian gauge potential interacts with non-Hermiticity in a realistic system, and what will happen to the corresponding topological phases~\cite{esaki2011edge,leykam2017edge}. To this end, we propose a method to realize a non-Abelian
	model with controllable lose and gain in a superconducting circuit. We find that: 
	(1) With only one of these two interactions, the non-Abelian gauge potential can lead to flying Hofstadter butterfly effect and 
	associated edge modes satisfying bulk-edge correspondence and the dissipation plays the role of loss and gain to each component. 
	(2) The non-commutation of the non-Abelian gauge potential and the non-Hermitian interaction can leads to coupling between these two terms, which can fundamentally influence the 
	fate of the topological phases. We find that with this non-Hermiticity, our model can host four different non-Hermitian phases: the gapped QSH phase, trivial gapped phase, gapless QSH phase and 
	metallic phase.
	(3) The bulk-edge correspondence is explicitly broken in the non-Hermitian model. By closing of the energy gap, the model can enter the gapless QSH regime, which also supports robust edge modes. The resonant couplings between the extended modes and localized edge modes are not allowed for their 
	different energies in the complex plane. 
	
	\begin{figure}
		\centering
		\includegraphics[width=1\linewidth]{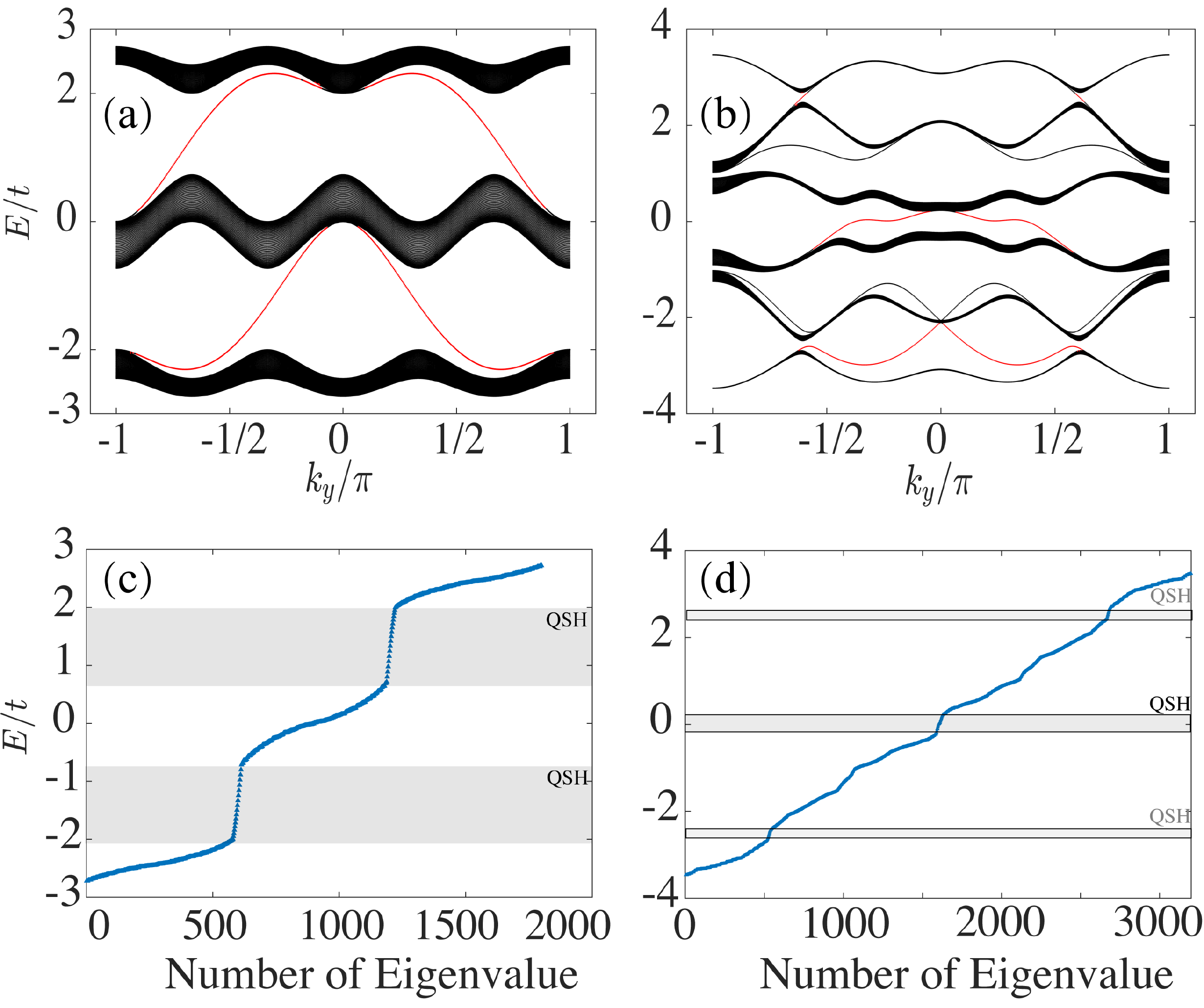}
		\caption{Edge modes in the non-Abelian gauge model without non-Hermiticity. The upper panel show the edge modes in a cylindrical geometry 
			for (a) $\gamma = 0$, $\lambda /t= 0$, $q = 3$ and $p = 1$; and (b) $\gamma = 1/4$, $\lambda t= 0.5$, $p = 1$ and $q = 6$. The edge modes in the
			gap are plotted using red thick lines. The lower panel show their corresponding edge modes in a finite system with (c) $30\times 30$ and 
			(d) $40\times 40$ with open boundaries in all directions.}
		\label{fig-fig2}
	\end{figure}
	
	{\it Model and Hamiltonian}. We consider the following tight-binding model in a square lattice (${\bf r} = (r_x, r_y)$, $r_{x,y} \in \mathbb{Z}$)
	\begin{equation}
	{\mathcal{H}_0} =  - \sum\limits_{{\bf r}, \sigma = x, y} t_\sigma \psi _{{\bf r} + {\bf e}_\sigma }^\dag U_{{\bf r}}^\sigma {\psi _{{\bf r}}} + \text{h.c.}, 
	\label{Eq:main_h}
	\end{equation}
	where ${\psi _{\r}} = {\left( {{a_{{\bf r}, \uparrow }}, {a_{{\bf r}, \downarrow}}} \right)^T}$ denotes the field operator of the (pseudo) spin component, $\sigma = \uparrow, \downarrow$, 
	and $t_\sigma$ is the nearest-neighbor hopping strength along directions ${\bf e}_x = (1,0)$ and ${\bf e}_y = (0,1)$. In this work we aim
	to explore a non-Abelian gauge potential with the form of $\mathcal{A} = - (2\pi \gamma \sigma_x, 2\pi \alpha x \sigma_z)$, which in the lattice model yields
	\begin{equation}
		\label{eq:hopping}
		U_{{\bf r}}^x= U^x = {e^{i{\Theta _x}}}, \quad U_{{\bf r}}^y=U^y_m=e^{i{\Theta _y}},
	\end{equation}
	with ${\Theta _x} = \gamma  \cdot 2\pi {\sigma _x}$, and ${\Theta _y} = \alpha  \cdot 2\pi m{\sigma _z}$. The same model has been studied in ultracold atoms 
	in Ref.~\cite{Osterloh2005}. This model can be realized using the lowest two modes in 3D cavities connected by superconducting quantum interference devices (SQUIDs),
	in which the energy mismatch during hopping can be compensated by the modulating frequencies of the driving SQUIDs; see our design details and simulations 
	in the supplementary material \cite{supp}.
	
	{\it Flying Hofstadter butterfly and edge modes}. Eq. \ref{Eq:main_h} can be used to explore the Hofstadter butterfly effect~\cite{hofstadter1976energy,azbel1964energy,claro1979magnetic,rammal1985landau}. For $\alpha=p/q$, where $p$, $q$ are coprimes, we can diagonalize the model in a reduced magnetic Brillouin zone (MBZ) \cite{supp}, yielding the following Harper equation,
	\begin{equation}
	\label{Eq:Haper}
	\begin{aligned}
	- {t_x}{{\tilde U}^{k_x^0 + 2\pi \alpha n}}{u_n} - {t_y}( {{{\tilde U}^{{k_y}}}{u_{n + 1}} + {{\tilde U}^{{-k_y}}}{u_{n - 1}}} )  = Eu_n,
	\end{aligned}
	\end{equation}	
	where ${{\tilde U}^{{k_x}}} ={e^{ - i{k_x}}}{U_x} + \text{h.c.}$, $\tilde U ^{k_y} = e^{-ik_y \sigma_z}$, and ${u_n} = {\left( {{u_{n, \uparrow }},{u_{n, \downarrow }}} \right)^T}$,
	with $u_{n, \sigma}$ being the amplitude of the wave function in the $n$th sector and $n = 1,2,\cdots, q-1$. The boundary condition is $u_{q} = {u_0}$ for $k_x^0 \in [\frac{-\pi}{q}, \frac{\pi}{q}]$ and 
	$k_y\in [-\pi,\pi]$. For a cylindrical geometry along $y$ direction (see Fig. ~\ref{fig-fig1} (a)), we have
	\begin{equation}
	\label{eq: ky}
	-t_x[ U^x u_{m+1} + {(U^x)^\dagger} u_{m-1}]-t_y(e^{ik_y}U^y_m + \text{h.c.})u_m = Eu_m.
	\end{equation}
	The butterfly diagram is presented in Fig.~\ref{fig-fig1} (b), which exhibits some features that are totally different from the Abelian case. 
	In the latter case, the flux parameterized by $k_y$ only slightly modify the edge modes~\cite{supp}; however, here it strongly influences the structure of the butterfly, exhibiting 
	some flying effect, that is, by tuning $k_y$ the butterfly pattern may also oscillate periodically~\cite{supp}. This arises from the coupling between the butterflies in each components, where 
	the non-Abelian gauge breaks the degeneracy at the crossing points. This butterfly structure has been studied by Osterloh {\it et al}~\cite{Osterloh2005} by fixing $k_y$ and tuning $\gamma$, 
	which do not have this flying feature. In our model the boundary condition can be modulated by threading a flux in the SQUID connecting the first and last sites, 
	which can be connected using SQUIDs. Thus this effect is observable in experiments.

	The intringuing things caused by the non-Abelian gauge field are the quantum spin Hall (QSH) effect and associated edge modes in a finite system. When $\gamma = 0$, the model is
	decoupled into two spinless copies with opposite magnetic field, which thus have opposite Chern numbers (see Fig.~\ref{fig-fig2}(a)). 
	In this case, the spin is conserved, and the spin Chern number
	can be well defined~\cite{sheng2006quantum, prodan2009robustness}. This QSH effect will be destroyed in the presence of $\gamma$ due to the closing of energy gap at $E = 0$ from
	the flying butterfly effect (see Fig.~\ref{fig-fig2} and  Ref.~\cite{supp}). To realize the QSH effect with finite $\gamma$, a staggered chemical potential should be introduced to open a gap, 
	which maybe written as
	\begin{equation}
	\label{Eq:Potential}
		\mathcal{V}_{\text{s}} = \sum_{\bf r} {\left( {-1} \right)^{r_x}}{\lambda}\psi _{{\bf r}}^\dag {\psi _{{\bf r}}}.
	\end{equation}
	This potential can be implemented in the SQUIDs with a finite frequencicy difference beween the 3D cavities and the modulated SQUIDs \cite{supp}. 

	To illustrate the QSH effect, in the following, we focus on $q = 6$ and $p = 1$. The numerical results are presented in Fig.~\ref{fig-fig2},
	which satisfy bulk-edge correspondence. In a cylindrical geometry, one can observe a number of edge modes in the gap, while in the open boundary condition, 
	these edge modes are localized at the two open ends within the energy gaps. In experiments, these modes separated from the extended bulk modes can be excited 
	and measured by the intensity of photons leaked from the edges by scanning of frequency. This technique has been implemented in experiments 
	for edge modes and topological insulator laser \cite{lu2015experimental,yang2018ideal, st2017lasing, harari2018topological}.

	{\it Non-Hermitian model and topological phases}.  The butterfly and edge modes provide important basis for us to explore the effect of non-Hermiticity on the topological phases. In the photonic 
	system, the loss and gain can be introduced to the model through interaction with the environment, which is described by the Lindblad superoperator. Let us define $\psi_{\bf r} = \text{Tr}(\rho c_{\bf r})$,
	where $\rho$ is the corresponding density matrix, we obtain the following Schr\"odinger equation~\cite{supp, Wang2016}, 
	\begin{equation}
	\label{eq:dynamic}
		id\Psi/dt = (\mathcal{M} -iK/2) \Psi,
	\end{equation}
	where $\Psi = (\psi_1, \psi_2, \cdots)^T$, $\mathcal{H}_0 = \sum_{{\bf r}, {\bf r'}} \psi_{\bf r}^\dagger (\mathcal{M})_{{\bf r}, {\bf r'}} \psi_{\bf r'}$, 
	and $K = 2\kappa \sigma_z$. The loss and gain in this scheme  can be controlled in the experiments \cite{Quijandria2018}. 
	Noticed that the above equation can be obtained from the Heisenberg equation $i\dot{\psi}_{\bf r} = [\psi_{\bf r}, \mathcal{H}]$, where $\mathcal{H} = \mathcal{H}_0 + \Psi^\dagger \kappa I\otimes \sigma_z \Psi. $After Fourier transformation in MBZ, we have \cite{supp},
	\begin{equation}
		H({\bf k}) = \Psi^\dagger (\tilde{U}^{k_y} \otimes S + V_s\otimes S^3 +  \text{h.c.}+V+ i\kappa \sigma_z\otimes {\bf{I}}) \Psi.
	\end{equation}
	In this equation $S$ is the cyclic permutation matrix with $S^q =1$, $V = \text{diag}(U^{k_x + 2\pi n/q})$, $n = 1,...,6$,and $\Psi = (\psi_1, \psi_2, \cdots, \psi_q)^T$. This model has several salient features: (\Roma{1}) It has a time-reversal symmetry $\mathcal{T} = i\sigma_y K$, where $K$ is
	the complex conjugate operator, whith $\mathcal{T} H({\bf k}) \mathcal{T}^{-1} = H(-{\bf k})$. Noticed that while a Zeeman field in the Hermitian model violate this symmetry, it does in the non-Hermitian model. 
	(\Roma{2}) It has a bi-chiral symmetry which is generalized from the chiral symmetry to
	\begin{equation}
	(\Gamma, H)_* = \Gamma H^\dagger + H \Gamma =0, (\Gamma, H)^* = H^\dagger\Gamma + \Gamma H =0.
	\end{equation}
	This symmetry ensures the eigenvalues $E_{n{\bf k}}$ and $-E_{n{\bf k}}^*$ appears in pairs. 
	(\Roma{3}) A new anti-unitary symmetry $Q = \sigma_x K$, with $Q H({\bf k}) Q^{-1} = H({\bf k})$. This symmetry ensures that 
	$E_{n{\bf k}}$ and $E_{n{\bf k}}^*$ come with pairs. In the Hermitian model, the bi-chiral relation is reduced to the 
	chiral relation, thus the combination of (\Roma{1}) and (\Roma{2}) can give rise to the charge-conjugate symmetry, making the system
	belongs to DIII class from ten-fold classification. In the presence of non-Hermiticity, the chiral symmetry is split to
	two bi-chiral relations, while the time-reversal symmetry is invariant. 

	An important feature is that the non-Hermitian term does not commute with the non-Abelian gauge potential when $\gamma \ne 0$, thus their interplay can give rise to some 
	intriguing physics. This consequence is trivial when $\gamma = 0$. In this case the two components only feel opposite but uniform dissipation and the eigenvectors 
	are unchanged while the eigenvalues will become complex, which only influence the intensity during time evolution.
	
	\begin{figure}
		\centering
		\includegraphics[width=0.9\linewidth]{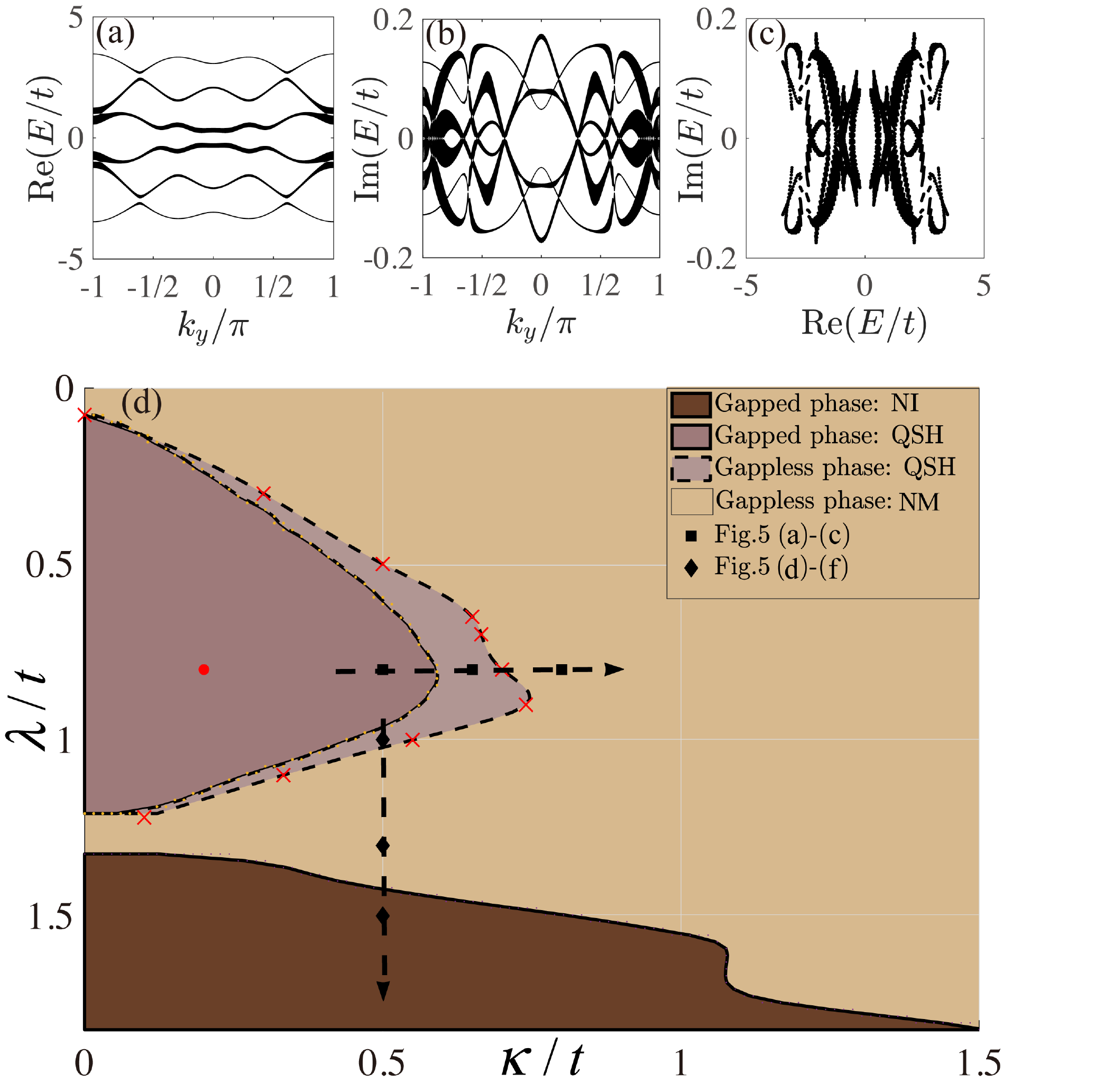}
		\caption{Phase diagram of non-Hermitian model with non-Abelian gauge potential with $\gamma = \frac{1}{4}$ and $\alpha = \frac{1}{6}$. (a) to (c) show the eigenvalues 
			for the red point in (d) with periodic boundary condition with $\kappa /t= 0.2$ and$ \lambda /t= 0.8$. The phase diagram in (d) is plotted by monitoring the gap closing of the bulk and associated 
			edge modes, which is divided into four different phases.  The edge modes for the solid symbols are presented in Fig. \ref{fig-fig4}.}
		\label{fig-fig3}
	\end{figure}
	
	\begin{figure}
		\centering
		\includegraphics[width=0.9\linewidth]{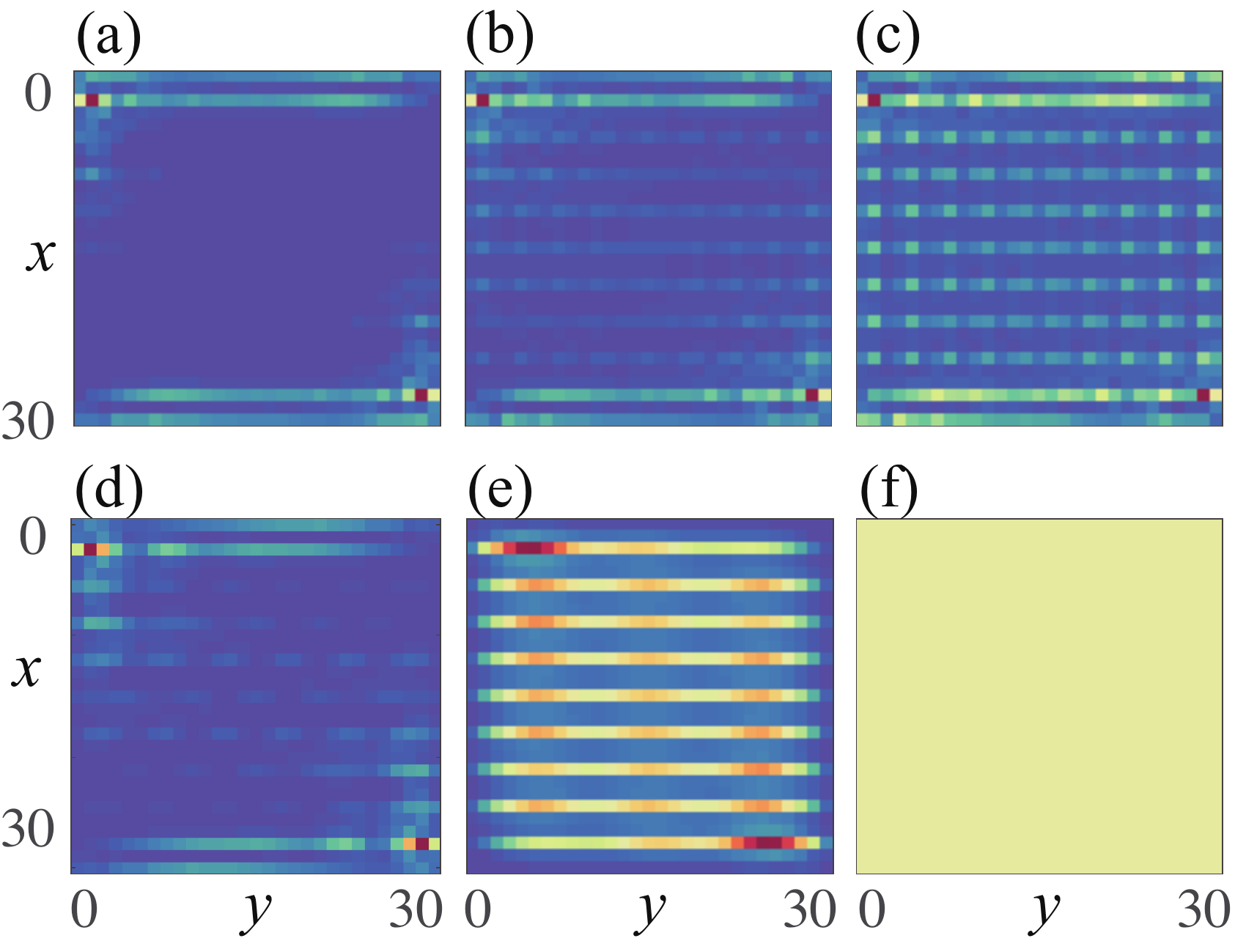}
		\caption{Edge modes in different phases with non-Hermiticity. The upper panel show the evolution for the three square symbols along the horizon line
 in Fig. \ref{fig-fig3} (d), thus (a) - (c) correspond to the edge modes in gapped QSH phase, gapless QSH phase and NM phase, respectively. Parameters are 
			$\lambda/t = 0.8$, $\kappa/t = 0.5$ (a), $0.65$ (b) and $0.8$ (3). 
			The lower panel show the phases for the three diamond symbols in Fig. \ref{fig-fig3} (d) with 
			$\kappa /t= 0.5$, $\lambda /t= 0.95$ (d), $1.3$ (e) and 1.5 (f), which correspond to the gapless QSH phase, NM phase and trivial gapped phase,
respectively. These results are demonstrated in a 30$\times$30 open lattice.}
	\label{fig-fig4}
	\end{figure}
	
	We solve the non-Hermitian model $\mathcal{H}$ in free space. In Fig.~\ref{fig-fig3}(a) and (b), we present the real and imaginary parts of the energy spectra, which show that 
	while the real part is fully gapped near $\text{Re}[E] \approx 0 $, the imaginary part can be gapless. By plotting these eigenvalues in the complex plane, 
	we find that the spectra form two separate groups, which can be separated from the whole complex plane. The symmetry of the eigenvalues about real and imaginary axis
	comes from the $Q$ symmetry and bi-chiral symmetry discussed above. This picture was used by Shen {\it et al}~\cite{Shen2018} to identify 
	the topological phases and associated edge modes, in which the edge modes play the role of connecting these two separated spectra. One should be noticed that the 
	topological non-Hermitian phases can only be well-defined in terms of symmetries, otherwise, one can construct a model without any symmetry as $\mathcal{H}' = e^{i\theta} \mathcal{H}$,
	which has the same eigenvalues in complex plane (upon a phase shift) and the same wave functions, including the edge modes, in open and infinite systems. 
	In our calculation, we identify the topological phases with non-Hermiticity by adiabatically switching off the loss and gain, and the non-Hermitian models 
	is said to have the same topology as the Hermitian ones if and only if that during adiabatic evolution (i) in the bulk spectra the two block of spectra are always disconnected 
	in periodic condition; and (ii) the edge modes connecting these two bulk blocks are always presented in open condition. In this sense, 
	the trivial gapped phase is characterized by absence of localized edge modes and respects criterion (i).

	Our phase diagram is presented in Fig. \ref{fig-fig3}. When $\kappa = 0$, we are able to identify three different phases: a normal insulator (NI) 
	phase, a gapped QSH phase and a normal metallic (NM) phase. These three phases can be understood from the discussions in Fig.~\ref{fig-fig2}, in which the phase 
	transition from the QSH phase to the NM phase satisfies bulk-edge correspondence. Moreover, the fully gapped phases can be characterized by $\mathbb{Z}_2$ indexes
	for DIII class, following Fukui {\it et al}~\cite{fukui2008topological}. 
	In the gapped QSH, we find $\nu = +1$; while in the trivial phase, $\nu = 0$. These three phases in the limit $\gamma = 0$ will help us to identify the phases with non-Hermiticity. 
	
	We are able to identify four different phases in the phase diagram by $\lambda$ and $\kappa$, three of which 
	can be connected to the well-defined phases in the Hermitian model mentioned above. Between the topological gapped QSH state and gapless NM phase, 
	we are able to identify a new phase denoted as gapless QSH state. We can understand this diagram by choosing a horizon line in Fig. \ref{fig-fig4} 
	with a fixed $0.07 \le \lambda/t \le 1.21$. By increasing of $\kappa$, the system may undergo two consequences, that is, the closing of energy gap from the fully 
	gapped phase to the gapless phase  at $\kappa_c^1$ in the bulk condition and the disappearance of edge state at $\kappa_c^2$ in the open system. We find that 
	$\kappa_c^1 < \kappa_c^2$, which is clear demonstration of the lack of bulk-edge correspondence, as unveiled in previous literature ~\cite{lee2016anomalous,
    zeuner2015observation,malzard2015topologically}. Thus the system may support three different phases, in which the new topological gapless phase between $\kappa_c^1 < \kappa < 
	\kappa_c^2$ can support edge modes in the gapless phase regime. 
	We present the evolution of edge modes in Fig. \ref{fig-fig4}, in which the two counter-propagating edge modes in the gapless QSH can be
	seen from Fig. \ref{fig-fig4} (b) and (d), for the two scanning lines in Fig. \ref{fig-fig3} (d). Noticed that in these results, we find that the edge modes are more likely to be localized 
	in the left-up and right-down corners due to the skin effect \cite{Yao2018, Yao2018a}, since in the plane wave basis, the momentum will become complex.

	In the Hermitian models, the gapless phase, in general, can not protect the robust edge modes due to the resonance coupling between the edge modes and 
	the bulk modes with the same energy. 
	This kind of resonant coupling can be forbidden in the non-Hermitian models, because although the total bulk spectra is connected in the complex plane, 
	the energy of the edge modes and the bulk modes can have different imaginary energies \cite{supp}. 
	For this reason, we are still able to find the robust edge modes in the gapless regime --- this is a quite general feature in all non-Hermitian topological models. In experiments, the 
	identification of these phases and associated counter-propagating edge modes can be detected by the external pumping at the edges \cite{supp}. 

	\textit{Conclusion} Topological non-Hermitian models have been widely explored based on spinless models. In future, their realizations with concrete materials should be 
	an important pursuit direction in the context of quantum simulation. Along this line, we generalize this idea to the realm of multi-component systems, and propose a general way for non-Hermitian 
	models with non-Abelian gauge potentials. These two non-commutative terms and their interplay can lead to various intriguing topological gapped and gapless phases, especially, we find 
	a topological gapless phase with robust edge modes due to the lack of bulk-edge correspondence. This platform can even be used to examine a number of important concepts in Hermitian models, 
	such as the butterfly spectra, in which non-Hermiticity can dramatically influence the fate of the Dirac cores~\cite{caibutterflyeffect}. We have also unveiled the consequence of time-reversal
	symmetry and chiral symmetry in terms of non-Hermiticity, which is important for the classification of Non-Hermitian topological phases.

	\textit{Acknowledgements.}  We thank Prof. Y.-H Wu and Prof. Z. Wang for helpful discussions. This project is supported by National Youth Thousand Talents Program (No. KJ2030000001), the USTC start-up funding (No. KY2030000053), the National Natural Science Foundation of China (No. 1177432, No. 11774114), the National Key Research and Development Program of China (No. 2016YFA0301700).
	
%	\bibliography{reference.bib}


\clearpage

\widetext

\begin{center}
	\textbf{\large Supplementary Material: Interplay between non-Hermiticity and non-Abelian gauge potential in topological photonics}
\end{center}
\setcounter{equation}{0}
\setcounter{figure}{0}
\setcounter{table}{0}
\setcounter{page}{1}
\titleformat{\section}{\centering\bfseries}{\Alph{section}.}{1em}{}`
%	\makeatletter
\renewcommand{\theequation}{S\arabic{equation}}
\renewcommand{\thefigure}{S\arabic{figure}}
\renewcommand{\thesection}{l\alph{section}}

\section{A. Physical Realization of the non-Abelian gauge potential} 

\begin{figure}
	\centering
	\includegraphics[width=0.9\textwidth]{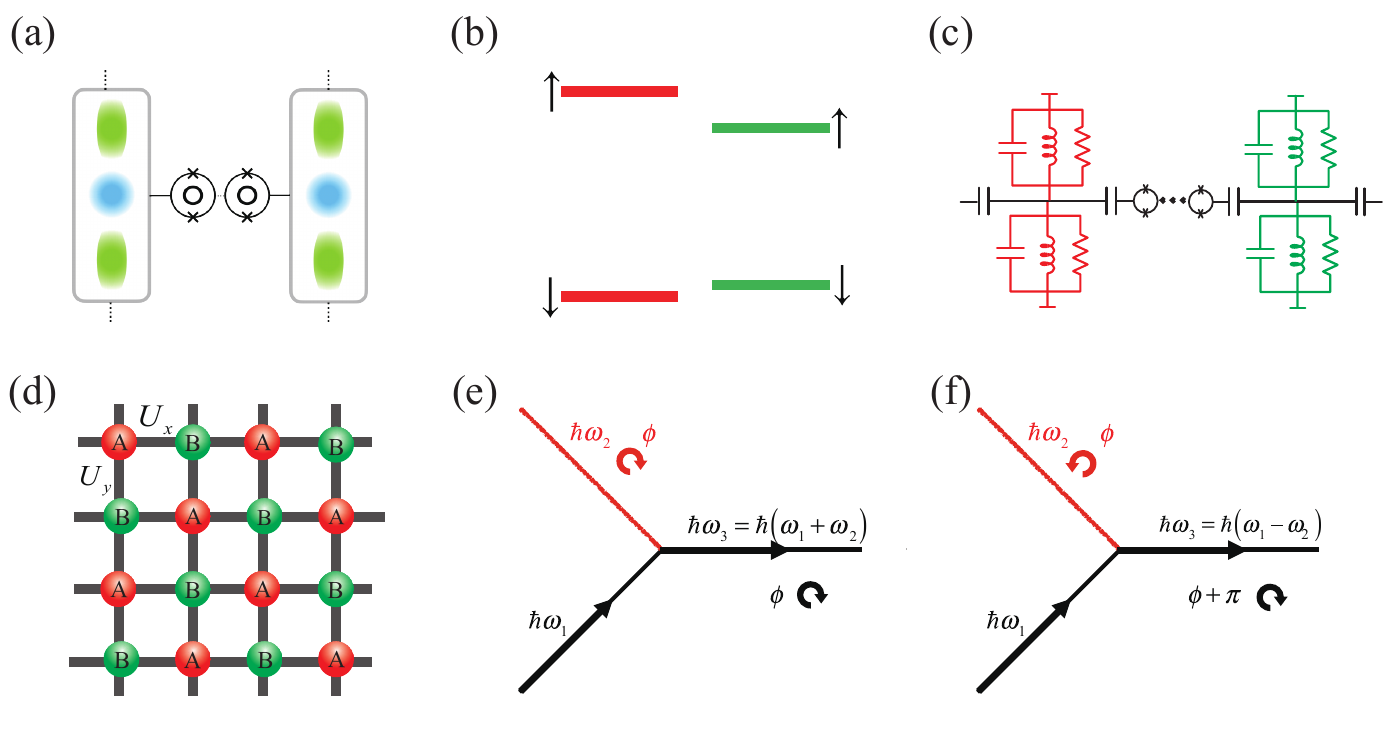}
	\caption{{\bf 3D cavity model and effective model}. (a) The structure of the 3D cavity $A$ (larger) and $B$ (smaller) used in this work. 
		(b) Energy level alignment of these two different cavities. (c) Effective circuit based on these two coupled cavities. 
		(d) Two dimensional 3D cavities connected in a square lattice for the realization of non-Abelian gauge potential and Non-Hermitian potential. 
		(e) - (f) Effective coupling mediated by the SQUIDs by periodical modulation. The rotating wave part and anti-rotating wave part from the 
		sum-frequency and difference-frequency parts can be controlled by the phase in the SQUIDs.}
	\label{fig-supp1}
\end{figure}

We discuss in details how to realize the non-Abelian gauge potential discussed in the main text. We consider the 3D cavities connected by
the superconducting quantum interference devices (SQUIDs) in a square lattice. In each cavity, we consider the lowest
two modes. Some typical experimental parameters are summarized in Table. 1, in which the energy level spacing between these two modes is typically of the order of GHz. The corresponding sizes of 
these cavity can be estimated as 
\begin{equation}
d = c/(2\nu) \sim  \text{ several centimeters}.
\end{equation}
For such a large size system, the uniformity of the 3D cavities can be made negligible. Let us denote these two energy levels as 
$\omega_{{\bf r},\sigma}$, where $\sigma = \uparrow$ and $\downarrow$. In the square lattice, we consider two different types of 3D cavities, which have different 
sizes (see Fig. \ref{fig-supp1} (a)). Thus by connecting these levels using SQUIDs, we obtain the following general Hamiltonian,
\begin{equation} 
H = \left(\sum\limits_{\substack{\left\langle {\r,\r'} \right\rangle \\ \sigma, \sigma' }} {{g_{\r,\r'}}\left( t \right)\psi _{\r',\sigma'}^\dag } {\psi _{\r,\sigma}} + \text{h.c.}\right) + \sum_{\r} \psi^\dagger_\r E_{\r} \psi_\r, 
\quad E_{\bf r} = \text{diag}(\omega_{{\bf r},\uparrow}, \omega_{{\bf r},\downarrow}) ,
\end{equation}
where in the first term the summation is carried out in the nearest-neighbor sites. The energy levels of these cavities are presented in Fig. \ref{fig-supp1} (b). The SQUID couples the four modes in the two
cavities, which is described by the following interaction 
\begin{equation}
g_{\r,\r'}\left(\tau\right) = \sum\limits_{\sigma ,\sigma '} {{t_{\r\sigma ,\r'\sigma '}}\cos \left( {{\Delta _{\r\sigma ,\r'\sigma '}}\tau + {\phi _{\r\sigma ,\r'\sigma '}}} \right)},
\end{equation}
where $\sigma ,\sigma ' \in \left\{ { \uparrow , \downarrow } \right\}$, $\tau$ denotes time and ${\Delta _{\r\sigma ,\r'\sigma '}}\left( \tau \right) = \left| {{\omega _{\r,\sigma }} - {\omega _{\r',\sigma '}}} \right|$. 
In this work we have assumed that these four levels have different energies, thus the tunneling between these four levels can be addressed individually. We design the cavities in such a way that 
the energy differences $\Delta_{ss'}$ should be large enough, thus during the fast modulation all the anti-rotating wave terms should be averaged out, giving rise to the following time-independent 
Hamiltonian
\begin{equation}
\label{Eq:coe}
\begin{aligned}
{\tilde H}_\text{eff} = \sum\limits_{\left\langle {\r,\r'} \right\rangle }{\psi _{\r'}^\dag {\Phi _{\r',\r}}{\psi _{\r}}} + h.c.  ,\quad 
{\left[ {{\Phi _{\r',\r}}} \right]_{\sigma ,\sigma '}} = \frac{{{t_{\r',\r,\sigma ,\sigma '}}{e^{-i { \phi _{\r',\r,\sigma ,\sigma '}}}}}}{2},
\end{aligned}
\end{equation}
where $\Phi_{\r',\r}$ is the hopping matrix hopping between lattice sites $\r$ and $\r '$. In this way, not only the coupling strength but also the relative phase carried by this tunneling 
can be controlled in experiments. We choose the following parameters:
\begin{equation}
\begin{aligned}
t_{\r, \r + e_x,\sigma,\sigma'} &= \left(\begin{array}{cc}
\cos(2\pi \gamma) & \sin(2\pi \gamma) \\ 
\sin(2\pi \gamma) & \cos(2\pi \gamma)
\end{array} \right)_{\sigma,\sigma'}, 
\quad t_{\r,\r+e_y} = \left(\begin{array}{cc}
1 & 0 \\ 
0 & 1
\end{array} \right)_{\sigma,\sigma'}, \\
\phi_{\r, \r+e_x,\sigma,\sigma'} &= -\left(\begin{array}{cc}
0 & \pi \\ 
\pi & 0
\end{array} \right)_{\sigma,\sigma'}, \quad 
\phi_{\r,\r+e_y,\sigma,\sigma' } = -\left(\begin{array}{cc}
2 \pi \r_x \alpha & 0 \\ 
0 & -2 \pi \r_x \alpha
\end{array} \right)_{\sigma,\sigma'}.
\end{aligned}
\end{equation}

Then we arrive at the hopping matrix in the main text as (assuming ${\bf r} = (m,n)$, with lattice constant $a = 1$)
\begin{equation}
U^x = \Phi_{\r,\r+e_x} = \exp(i2\pi \gamma \sigma_x), \quad 
U^y(x) = \Phi_{\r,\r+e_y}  = \exp(i 2\pi m \sigma_z).
\end{equation}

This model provides an useful toolbox to simulate the other interesting physics. By carefully engineering these parameters, one can in principle realize arbitrary types of 
non-Abelian gauge potentials. For instance, the widely explored Rashba spin-orbit coupling can also be simulated in this model using the following parameters,
\begin{equation}
\begin{aligned}
t_{\r, \r + e_x,\sigma,\sigma'} &= \left(\begin{array}{cc}
\cos(2\pi \gamma) & \sin(2\pi \gamma) \\ 
\sin(2\pi \gamma) & \cos(2\pi \gamma)
\end{array} \right)_{\sigma,\sigma'}, 
\quad t_{\r,\r+e_y} = \left(\begin{array}{cc}
\cos(2\pi \beta) & \sin(2\pi \beta) \\ 
-\sin (2\pi \beta) & \cos(2\pi \beta)
\end{array} \right)_{\sigma,\sigma'},\\
\phi_{\r, \r+e_x,\sigma,\sigma'} &= -\left(\begin{array}{cc}
0 & \pi \\ 
\pi & 0
\end{array} \right)_{\sigma,\sigma'}, \quad \phi_{\r,\r+e_y,\sigma,\sigma' } = -\left(\begin{array}{cc}
0 & 0 \\ 
0 & 0
\end{array} \right)_{\sigma,\sigma'},
\end{aligned}
\end{equation}
in which the hopping matrix can be written as
\begin{equation}
U^x = \exp(i 2 \pi \gamma \sigma_x) \quad U^y = \exp(i 2 \pi \beta \sigma_y).
\end{equation}

This idea may be generalized to multicomponent conditions. This kind of tunability is still challenging in other physical systems. Finally, we need to emphasize that the long range 
interaction can also be realized in this model by connecting two far-distant 3D cavities by the SQUIDs. 

\vspace{0.3cm}

To realize the staggered potential used in Eq. 6 in the main text for the realization of gapped QSH, one can choose the following parameters:
\begin{equation}
{\Delta _{\r\sigma ,\r'\sigma '}} = \left| {{\omega _{\r,\sigma }} - {\omega _{\r',\sigma '}}} \right| - (-1)^{r_x-r'_x} 2\lambda ,
\end{equation}
where $r_x$ is the site index in $x$-direction. If we define ${\bf r} = (m,n)$ and ${\bf r}' = (m',n')$, then $r_x = m-m'$. 

\section{B. Simulation of the effective Hamiltonian}

To gain a much better understanding of the effective Hamiltonian used in the main text, we discuss the effective time-independent Hamiltonian derived from the modulating 
Hamiltonian. We consider a two-site model with the following Hamiltonian,
\begin{equation}
H_\text{two-sites} = \sum_{\sigma = \uparrow / \downarrow , i = 1,2}\omega_{\sigma,i} n_{\sigma,i} + 2 t \sum_{\sigma, \sigma'} \cos(\Delta_{\sigma 1,\sigma' 2} \tau + \phi_{\sigma 1,\sigma' 2}) 
(a^\dagger_{\sigma,1} a_{\sigma',2}+h.c.) ,
\end{equation}
where the evolution of time is denoted by $\tau$. We can choose parameters
\begin{equation}
\omega_{\downarrow,1} = 8.7 \times 2\pi  ~\text{GHz}, \quad \omega_{\uparrow,1} = 9.3 \times 2\pi ~\text{GHz}, \quad  \omega_{\downarrow,2} = 9.1 \times 2\pi ~\text{GHz}, \quad 
\omega_{\uparrow,1} = 9.9 \times 2\pi ~\text{GHz}.
\end{equation}
The experimentally feasible parameters for the hopping strengths are
\begin{equation}
2t = 2.4 ~\text{MHz}, \quad 2 \lambda = 5 ~\text{MHz}.
\end{equation}

In the above chosen parameters, the energy difference between these two frequencies are much large than $2t$ and $2 \lambda$, thus in the following, we will use the 
rotating wave approximation to eliminate the fast oscillation terms, leaving a static Hamiltonian. 

\begin{table}
	\centering
	\begin{adjustbox}{width=0.8\textwidth}
		\small
		\begin{tabular}{|l|l|l|l|l|}
			\hline 
			Ref. & Modes & Frequency (GHz $\times 2\pi$) & $Q_{\text{int}}$ & $T_1$ \\ 
			\hline 
			\cite{Paik2011}& TE101,TE102 & 8--10~GHz & $2\times 10^6, 5\times 10^5$ & $50$~$\mu \text{s}$ \\ 
			\hline 
			\cite{reagor2013reaching}& TE101,TE111,TE011,TM111 & 7.6--11.5 & $10^6, 10^8$ & $0.01 \text{ms} $--$ 10.4$~$ \text{ms}$ \\ 
			\hline 
			\cite{Sirois2015}& TE101 & 8.7--9.33 & $0.8\times 10^6, 10^8$ & $14.9$~$\mu\text{s}$--$0.93$~$\mu\text{s}$ \\ 
			\hline 
			\cite{reagor2016quantum}& TM01 & 9.8 & $2\times 10^8$ & $0.72\text{ms}$--$7$~$\text{ms}$ \\ 
			\hline 
			\cite{owens2018quarter}& / & 9.560 & $1500-3000$ & / \\ 
			\hline 
		\end{tabular} 
		\label{Table:1}
	\end{adjustbox}
	\caption{{\bf Possible parameters in recent 3D cavity experiments}. In different literature, both TE modes and TM modes are used in experiments, depending strongly on 
		the geometric sizes of the 3D cavities. The typical frequencies are of the order of GHz, with $Q$ factor from $10^6$ to $10^8$. The last column shows the lifetime 
		of the energy levels. }
\end{table}

\begin{figure}
	\centering
	\includegraphics[width=0.99\linewidth]{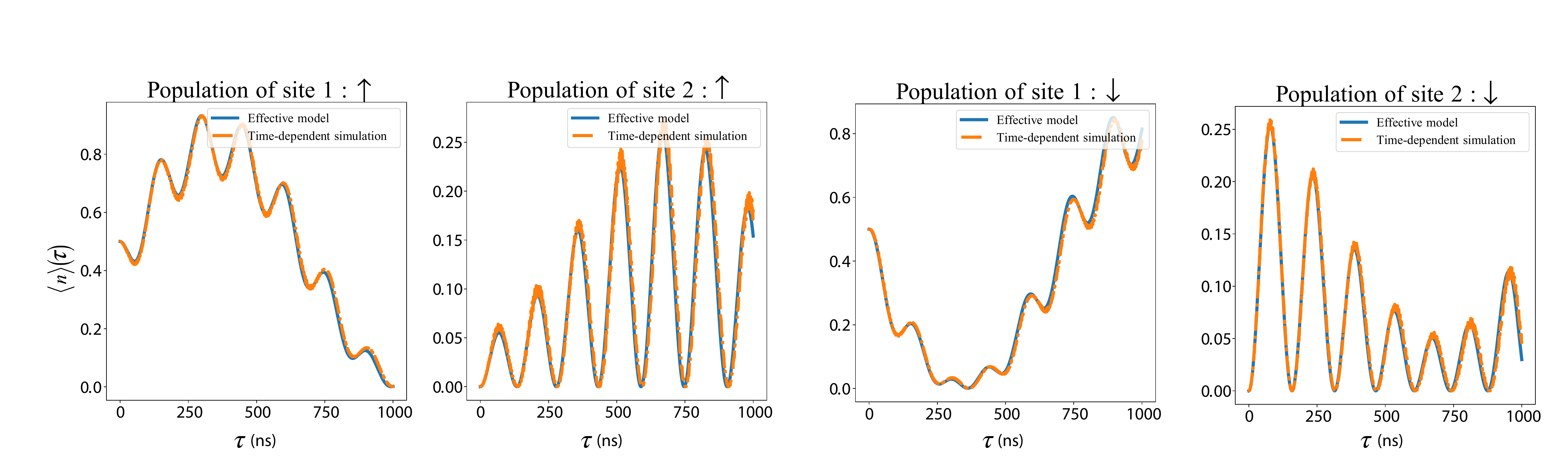}
	\caption{{\bf Dynamics from exact Hamiltonian and effective Hamiltonian}. We calculate the mean population 
		in the time-dependent Hamiltonian and the effective Hamiltonian, start from the same initial wave function. We choose 
		$\phi_{1,2} = \left(\begin{array}{cc}
		\pi/\sqrt{7} & \pi/\sqrt{2} \\ 
		\pi/\sqrt{3} & \pi/\sqrt{5}
		\end{array} \right)$ for demonstration. }
	\label{fig-supp2}
\end{figure}

In the rotating frame let us define 
\begin{equation}
U = \exp \left\{-i \tau \left[ \sum_{\sigma = \uparrow / \downarrow , i = 1,2}\omega_{\sigma,i} n_{\sigma,i} -2 \lambda(n_{\uparrow,1} + n_{\downarrow,1}) \right]\right\}, 
\end{equation}
then we have following effective Hamiltonian,
\begin{equation}
\begin{aligned}
\tilde{H} &= U^\dagger H_{\text{two-sites}} U + i (\partial_\tau U^\dagger )U \\
&= 2 \lambda (n_{\uparrow,1} + n_{\downarrow,1}) +t \sum_{\sigma, \sigma'} \left[e^{i(\Delta_{\sigma 1,\sigma' 2} \tau + \tilde \phi_{\sigma 1, \sigma' 2})} + \text{h.c.} \right]
\times\left[ e^{i (\omega_{\sigma,1}-\omega_{\sigma,2} - 2\lambda)\tau} a_{\sigma,1} a^\dagger_{{\sigma',2}} + \text{h.c.}\right],
\end{aligned}
\end{equation}
where $\tilde \phi _{\r',\r,\sigma ,\sigma '} = \text{sign} (\omega_{\r',\sigma '} - \omega_{\r,\sigma }) \phi _{\r',\r,\sigma ,\sigma '}$. In the above equation, the fast oscillation with
period much short than the energy scale of $t$ and $\lambda$ are neglected, leaving the following effective Hamiltonian
\begin{equation}
H_{\text{eff}} = \lambda n_1  - \lambda n_2 +  t \psi^\dagger_1 \Phi_{1,2} \psi_2 + \text{h.c.}).
\end{equation}
We examine the validity of the above effective Hamiltonian by checking the dynamics of the particle in this two site model, which is shown in Fig. \ref{fig-supp2}. 
We find that these two Hamiltonians ($H_\text{two-sites}$ and $H_{\text{eff}}$) will give the same dynamics. The same feature can be found for the other physical quantities, 
independent of its initial condition. Thus the effective Hamiltonian can faithfully describe the dynamics of the time-dependent Hamiltonian, which can be generalized 
to the whole lattice systems, as was used in the main text.

\section{C. Engineering of the Dissipation}

In this section we consider the following more general model,
\begin{equation}
H = \sum_{{\bf r},{\bf r'}} t_{{\bf r},{\bf r'}} a_{\bf r}^\dagger a_{\bf r'},
\end{equation}
which is subjected to the Lindblad master equation, 
\begin{equation}
i\dot{\rho} = [H,\rho] + \underbrace{i \sum_\alpha \mathcal{L}_\alpha \rho \mathcal{L}_\alpha^\dagger - {i \over 2} \{\mathcal{L}_\alpha^\dagger \mathcal{L}_\alpha, \rho\}}_{\mathcal{L}_{\text{d}}} .
\end{equation}

When $\mathcal{L}_\alpha$ is the dissipation operator. We consider two elementary dissipations with $L^+_{\r} = \sqrt{\gamma^+} a_{\r}$ and $L^-_{\r} = 
\sqrt{\gamma^-} a_{\r}$, which correspond to particle loss and particle injection, respectively. To obtain the evolution of the coherent state, one can simply 
take the average of the field operator, i.e., 
\begin{equation}
\label{eq: Supp8}
i\frac{d\left\langle a_{\r}\right\rangle }{dt}  = i \text{Tr}[a_\r \dot{\rho}] = \text{Tr} [ a_{\r}H\rho - a_{\r}\rho H + i a_{\r} \mathcal{L}_{\text{d}}]
\end{equation}
It is easy to obtain that 
\begin{equation}
\text{Tr} [a_{\r}H\rho - a_{\r}\rho H] = \text{Tr} (\rho [a_{\r}, H]) = \sum_{\bf r'} t_{{\bf r},{\bf r'}} \langle a_{\bf r'} \rangle,
\end{equation}
which is true for both fermions and bosons. For the dissipation term related to $\mathcal{L}_d$, we have the following two identities,
\begin{equation}
\begin{aligned}
&& \text{Tr}\sum\limits_{\r'} {\left[ {{a_\r}{a_{\r'}}\rho {a_{\r'}}^\dag  - \frac{1}{2}{a_\r}a_{\r'}^\dag {a_{\r'}}\rho  - \frac{1}{2}{a_\r}\rho a_{\r'}^\dag {a_{\r'}}} \right]} = \frac{1}{2}\text{Tr}\sum\limits_{\r'} {\left[ {{a_{\r'}}^\dag {a_\r}{a_{\r'}}\rho  - {a_\r}a_{\r'}^\dag {a_{\r'}}\rho } \right]}  =  - \frac{1}{2} \left\langle a_\r \right\rangle,\\
&& \text{Tr}\sum\limits_{\r'} {\left[ {{a_\r}{a_{\r'}}^\dag \rho {a_{\r'}} - \frac{1}{2}{a_\r}{a_{\r'}}a_{\r'}^\dag \rho  - \frac{1}{2}{a_\r}\rho {a_{\r'}}a_{\r'}^\dag } \right]} = \frac{1}{2}\text{Tr}\sum\limits_{\r'} {\left[ {{a_{\r'}}{a_\r}{a_{\r'}}^\dag \rho  - {a_{\r'}}a_{\r'}^\dag {a_\r}\rho } \right]}  = \frac{1}{2} \left\langle a_\r \right\rangle.
\end{aligned}
\end{equation}
Thus if one defines $\psi_{\bf r} = \text{Tr}(\rho a_{\bf r})$, then the dynamics of $\Psi = (\psi_1, \psi_2, \psi_3, \cdots)$ can be described by the following non-Hermitian Schr\"odinger equation,
\begin{equation}
i\partial_t \Psi = (H + i\Gamma) \Psi,
\end{equation}
where for the on-site dissipation studied above, we have
\begin{equation}
\Gamma = \text{diag}(\gamma_1, \gamma_2, \gamma_3, \cdots).
\end{equation}

A few remarks about these results are in orders. (1) When $\Gamma = 0$ with conserved total number of particle, the trace $\text{Tr}(\rho a_{\bf r}) = 0$ exactly, thus the above equation is null. 
It is a nontrivial equation which can be used to describe dynamics of quantum state (in the single particle level) only when dissipation and relaxation are considered. 
(2) This equation has several interesting limits. In one hand when all $\gamma_i$ have the same value, it only introduces a global dissipation to all quantum state, since $[\Gamma, H] = 0$, thus the 
effect of dissipation is somewhat trivial \cite{lu2018spontaneous}. (3) This dissipation can be used to explore some interesting physics such as models with $\mathcal{PT}$-symmetry 
\cite{Quijandria2018}. (4) In this work, we aim to explore the interesting physics due to interplay
between non-Hermicity and non-Abelian gauge potential with $[H, \Gamma] \ne 0$, which can give rise to new physics in our work. 

\section{D. Band Structure in momentum Space}

\begin{figure}
	\centering
	\includegraphics[width=0.9\linewidth]{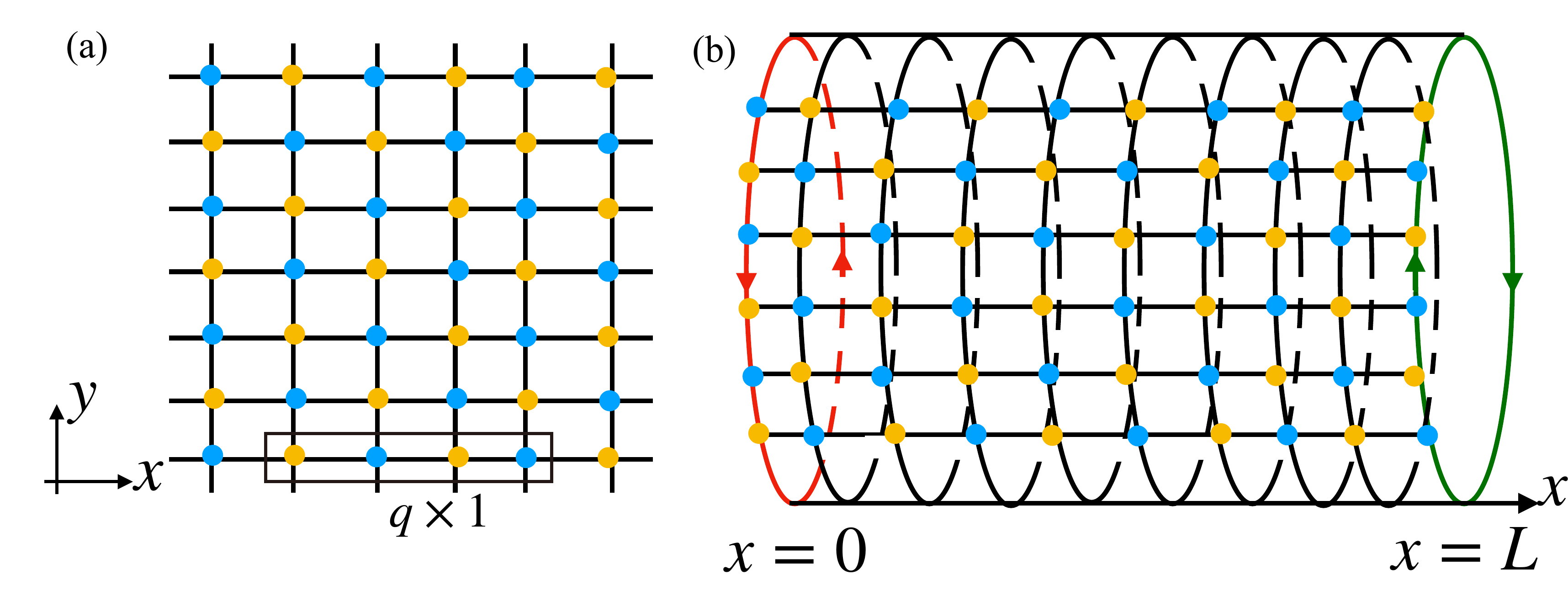}
	\caption{{\bf Two different geometries in our calculation}. (a) Infinite square lattice. The non-Abelian gauge will enlarge the real space periodicity  by $q$ times,
		where $\alpha = p/q$. (b) The cylindrical geometry, in which two localized states can be found localized in the edge.}
	\label{fig-supp3}
\end{figure}

In this section, we discuss the calculation of the energy bands from Bloch theorem. We focus on two geometries as shown in Fig. \ref{fig-supp3} (a) 
in infinite size system and (b) in a cylindrical geometry. We take the Hamiltonian defined in Eq. 1 and the gauge field in Eq. 2 in the main text as an example 
to show the calculation of energy bands in momentum space. This model may also be formally written as $H = -t_x {T_x} -t_y {T_y} + + V_{\text{stag}}$. 

Firstly, we make the following Fourier transform to the field operators by
\begin{equation}
{\psi _{m,n}} = \int_{ - \pi }^\pi  {\frac{{d{k_x}}}{{2\pi }}\int_{ - \pi }^\pi  {\frac{{d{k_y}}}{{2\pi }}{e^{i{k_x}m + i{k_y}n}}{\psi _{{k_x},{k_y}}}} } ,
\end{equation}
where the momenta $k_x$ and $k_y$ are constraint in the first magnetic Brillouin zone (MBZ) $\left[ { - \pi ,\pi } \right]$. 
The hopping term $T_x$ along $x$ direction is homogeneous, leading to a form in $k$-space as
\begin{equation}
\begin{gathered}
{T_x} = \sum\limits_{m,n} {\psi _{m + 1,n}^\dag {U^x}{\psi _{m,n}} + h.c.}  = \int_{ - \pi }^\pi  {\frac{{d{k_x}}}{{2\pi }}\int_{ - \pi }^\pi  {\frac{{d{k_y}}}{{2\pi }}\psi _{{k_x},{k_y}}^\dag } } {{\tilde U}^{{k_x}}}{\psi _{{k_x},{k_y}}},
\end{gathered} 
\end{equation}
where
\begin{equation}
{{\tilde U}^{{k_x}}} ={e^{ - i{k_x}}}{U_x} + h.c. = 2   \left( {\begin{array}{*{20}{c}}
	{\cos {k_x}\cos 2\pi \gamma }&{\sin {k_x}\sin 2\pi \gamma } \\ 
	{\sin {k_x}\sin 2\pi \gamma }&{\cos {k_x}\cos 2\pi \gamma } 
	\end{array}} \right).
\end{equation}
The hopping  term $T_y$ along $y$ direction is not diagonal in $k$-space, and after Fourier transformation we have
\begin{equation}
\begin{aligned}
{T_y} &= \sum\limits_{m,n} {\psi _{m,n + 1}^\dag U_m^y{\psi _{m,n}} + h.c.}  \\
&= \int_{ - \pi }^\pi  {\frac{{d{k_x}}}{{2\pi }}\int_{ - \pi }^\pi  {\frac{{d{k_y}}}{{2\pi }}\psi _{{k_x} + 2\pi \alpha ,{k_y}}^\dag } } {{\tilde U}^{{k_y}}}{\psi _{{k_x},{k_y}}} + \int_{ - \pi }^\pi  {\frac{{d{k_x}}}{{2\pi }}\int_{ - \pi }^\pi  {\frac{{d{k_y}}}{{2\pi }}\psi _{{k_x} - 2\pi \alpha ,{k_y}}^\dag } } {{\tilde U}^{ - {k_y}}}{\psi _{{k_x},{k_y}}},\\ 
\end{aligned}
\end{equation}
where $\tilde U ^{k_y} = e^{-ik_y \sigma_z}$.

The gauge potential can enlarge the MBZ in real space (see Fig. \ref{fig-supp3} (a)), in which in our model the MBZ is made by $q\times 1$ sites. This increased MBZ in real space will shrink the size of the
reciprocal lattice in momentum space. We can write the total Hamiltonian in momentum space as following
\begin{equation}
H = \int_{ - \pi /q}^{\pi /q} {\frac{{dk_x^0}}{{2\pi }}} \int_{ - \pi }^\pi  {\frac{{d{k_y}}}{{2\pi }}} \sum\limits_{n = 0}^{q - 1} {{H_{k_x^0,{k_y},n}}},
\end{equation}
where ${{H_{k_x^0,{k_y},n}}}$ reads as
\begin{equation}
\begin{aligned}
{H_{k_x^0,{k_y},n}} &=  - {t_x}\left( {\psi _{k_x^0 + 2\pi \alpha n}^\dag {{\tilde U}^{{k_x} + 2\pi \alpha n}}{\psi _{k_x^0 + 2\pi \alpha n,ky}}} \right)- {t_y}\left( {\psi _{k_x^0 + 2\pi \alpha \left( {n + 1} \right)}^\dag {{\tilde U}^{{k_y}}}{\psi _{k_x^0 + 2\pi \alpha n,ky}}} \right)\\
&- {t_y}\left( {\psi _{k_x^0 + 2\pi \alpha \left( {n - 1} \right)}^\dag {{\tilde U}^{ - {k_y}}}{\psi _{k_x^0 + 2\pi \alpha n,ky}}} \right). \\ 
\end{aligned}
\end{equation}

This model in momentum space can be diagonalized using the following basis,
\begin{equation}
\left| {{\Psi _{ \uparrow / \downarrow }}} \right\rangle  = \sum\limits_{n = 0}^{q - 1} {{u_{n, \uparrow / \downarrow }}\psi _{k_x^0 + 2\pi \alpha n,{k_y}, \uparrow / \downarrow }^\dag \left| 0 \right\rangle}, 
\end{equation}
where ${u_n} = {\left( {{u_{n, \uparrow }},{u_{n, \downarrow }}} \right)^T}$ is the amplitude of the relative wave function. The secular equation with these coefficients can be written as 
\begin{equation}
\begin{gathered}
\begin{aligned}
- {t_x}{{\tilde U}^{k_x^0 + 2\pi \alpha n}}{u_n} - {t_y}\left( {{{\tilde U}^{{k_y}}}{u_{n + 1}} + {{\tilde U}^{{-k_y}}}{u_{n - 1}}} \right)= E\left( {k_x^0,{k_y}} \right){u_n},
\end{aligned}
\end{gathered} 
\end{equation}
subject to the boundary condition as 
\begin{equation}
u_{n + q} = u_n,  \quad n = 0, 1, 2, 3, \cdots, q -1.
\end{equation}

The staggered potential $V_{\text{stag}}$ is essential to realize the topological gapped QSH state, which can be written as
\begin{equation}
\begin{aligned}
{V_{stag}/\lambda} &= \sum\limits_{m,n} {{{\left( { - 1} \right)}^m}\psi _{m,n}^\dag {\psi _{m,n}}}  \\
&= \int_{ - \pi }^\pi  {\frac{{d{k_x}}}{{2\pi }}\int_{ - \pi }^\pi  {\frac{{d{k_y}}}{{2\pi }}} } \psi _{{k_x} + \pi ,{k_y}}^\dag {\psi _{{k_x},{k_y}}} 
= \int_{ - \pi /q}^{\pi /q} {\frac{{d{k_x}}}{{2\pi }}\int_{ - \pi }^\pi  {\frac{{d{k_y}}}{{2\pi }}} } \sum\limits_n {\psi _{k_k^0 + \pi  + 2\pi \alpha n,{k_y}}^\dag {\psi _{k_k^0 + 2\pi \alpha n,{k_y}}}}. 
\end{aligned}
\end{equation}
When $q$ is even, we have $k_x^0 + \pi  + 2\pi \alpha n = k_x^0 + 2 \pi \alpha(n + \frac{q}{2}) - 2\pi (p-1)$. Hence this potential realizes coupling between the $n$-component and 
$n + q/2$-component, and opens a finite gap. From the symmetry perspective, this term breaks the axial symmetry which make the original degenerate bands further split into more non-degenerate bands.

We can also take the cylindrical version of this model to examine the effect of edge modes. Around the cylindrical geometry, the momentum is a good quantum number and we have
\begin{equation}
H = \int {\frac{{d{k_y}}}{{2\pi }}\left( {\sum\limits_m {\psi _{m,{k_y}}^\dag {U_{on}}\left( {m,{k_y}} \right){\psi _{m,{k_y}}}}  + h.c.} \right)}  + \sum\limits_m {\psi _{m + 1,{k_y}}^\dag {U_x}\left( m \right){\psi _{m,{k_y}}}},
\end{equation}
where \begin{equation}
\begin{aligned}
{U_{on}}\left( {m,{k_y}} \right) &= \left( {\begin{array}{*{20}{c}}
	{ - 2{t_y}\cos \left( {{k_y} - 2\pi m\alpha } \right) + {{\left( { - 1} \right)}^m}{\lambda _{stag}}}&0\\ 
	0&{ - 2{t_y}\cos \left( {{k_y} + 2\pi m\alpha } \right) + {{\left( { - 1} \right)}^m}{\lambda _{stag}}} 
	\end{array}} \right),\\
{U_x} &= \left( {\begin{array}{*{20}{c}}
	{\cos \left( {2\pi \gamma } \right)}&{i\sin \left( {2\pi \gamma } \right)} \\ 
	{i\sin \left( {2\pi \gamma } \right)}&{\cos \left( {2\pi \gamma } \right)} 
	\end{array}} \right).
\end{aligned}
\end{equation}
This method is used to check the edge excitations in the gap. 

\section{E. Symmetries of the Hamiltonian in momentum space}

In this section, we will illustrate the symmetries in the Harper equation step by step. Firstly, we consider 
the spinless model without dissipation, which has been studied in previous literature. Let us assume the hopping amplitude 
$t_x, t_y =1$, then we have
\begin{equation}
H({\bf k}) = H(k_x, k_y) = e^{-ik_y} S + e^{-ik_x} V + \text{h.c.},
\label{eq-Hk}
\end{equation}
where $p$ and $q$ are coprimes, $S$ is the cyclic permutation matrix and $V$ is a diagonal matrix with element entries $\{e^{- i\Phi n}\}, n = 1,2, \cdots, q$ 
and $\Phi = 2\pi p/q$. For example with $q = 4$, we can write down these two matrices explicitly as \cite{wen1989winding}
\begin{equation}
S  = \left[ \begin{array}{cccc}
& 1 &  &  \\ 
&  & 1 &  \\ 
&  &  & 1 \\ 
1 &  &  & 
\end{array} \right], \quad V = \left[\begin{array}{cccc}
e^{-i \Phi \times 1} &  &  &  \\ 
& e^{- i \Phi \times 2} &  &  \\ 
&  & e^{-i \Phi \times 3} &  \\ 
&  &  & e^{- i \Phi \times 4}
\end{array} \right].
\end{equation}	
The similar forms can be generalized to arbitrary $q$. We can check that
\begin{equation}
S^q = 1, \quad  SV = VS e^{i\Phi}.
\end{equation}
Moreover, under the transformation of these matrices, we have 
\begin{equation}
VH(k_x,k_y)V^{-1} = H(k_x + \Phi, k_y), \quad SH(k_x,k_y)S^{-1}=H(k_x,k_y+\Phi).
\end{equation}

\begin{figure}
	\centering
	\includegraphics[width=0.7\textwidth]{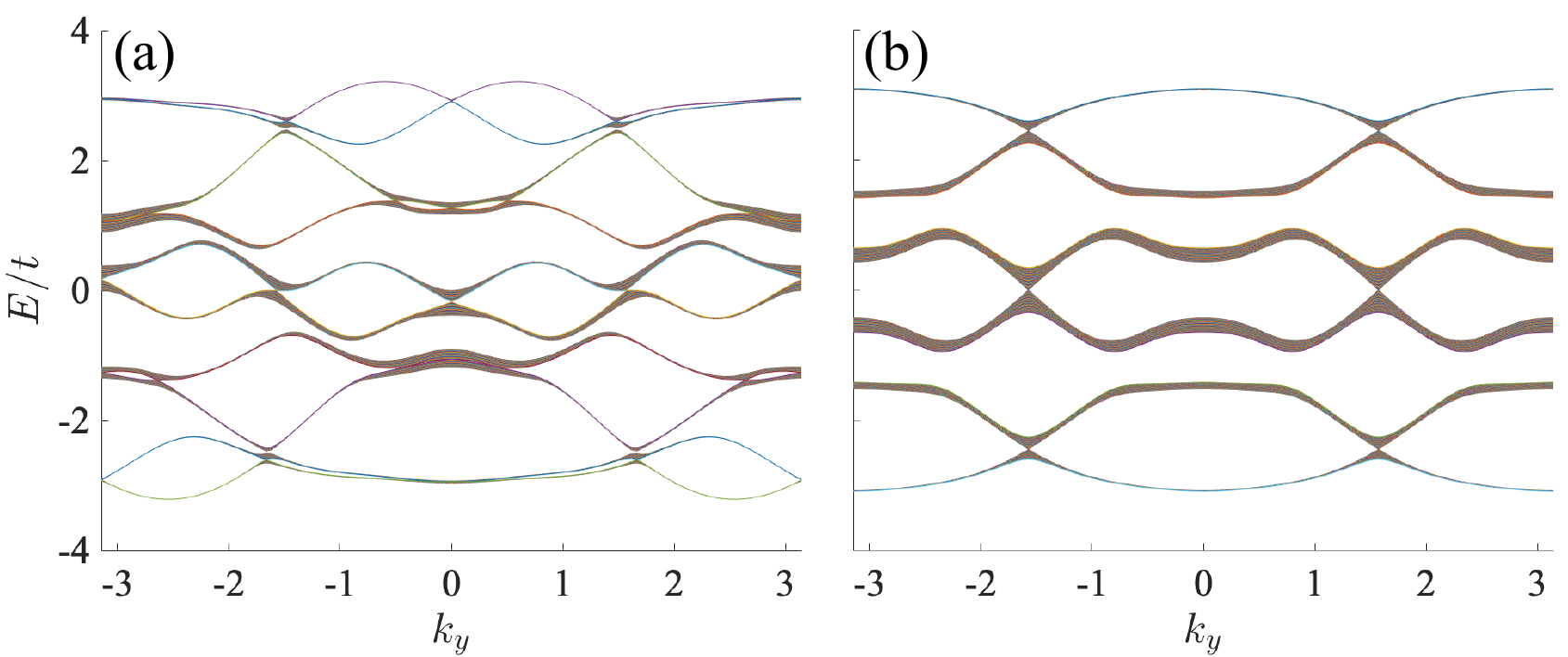}
	\caption{{\bf Band structures in the Harper equation}. (a) $p=1$, $q= 5$ and (b) $p=1$, $q= 6$. The even and odd cases will have totally different symmetries. 
		In this work, we focus on $q$ to be an even number, in which the eigenvalues $E$ and $-E$ will appear in pairs. This result is related to the chiral symmetry
		defined in this supplementary material.}
	\label{fig-supp4}
\end{figure}

The band structures for $p=1$, $q = 5$ and $q = 6$ are presented in Fig. \ref{fig-supp4}. We find that when $q$ is an even number, the spectra is always 
symmetric about $E = 0$, which is a general feature not only in this spinless model, but also exists in the models we have considered in the main text. This symmetry
is related to the chiral operator $\Gamma$. For $q = 4$, this operator can be written as
\begin{equation}
\Gamma = \left[\begin{array}{cccc}
0 & 0 & 1 & 0 \\ 
0 & 0 & 0 & -1 \\ 
1 & 0 & 0 & 0 \\ 
0 & -1 & 0 & 0
\end{array}\right]. 
\end{equation}

This operator can be generalized to the other Harper equations when $q$ is even. If we define the eigenvectors of the above model 
as $|u_j\rangle$, then we have
\begin{equation}
\langle u_j | \Gamma | u_{j+q/2}\rangle = \delta_{j,j+q/2} (-1)^j (i)^{q/2},
\label{eq-Gamma}
\end{equation}
after a proper choose of the global phase. It's easily to check that 
\begin{equation}
\left\{\Gamma,H \right\} = \left\{\Gamma,V \right\} = \left\{\Gamma,S \right\} = 0, \quad \Gamma^2 = {\bf I}.
\end{equation}
where we denote ${\bf I}$ to be the unity matrix. This symmetry ensures the appearance of $E_n$ and $-E_n$ pairs in the spectra. 
When $H\phi_n = E_n \phi_n$, then $H\Gamma \phi_n = -\Gamma H \phi_n = -\Gamma E_n \phi_n$, thus $\Gamma \phi_n$ is the eigenstate of the 
Hamiltonian with eigenvalue $-E_n$. Noticed that in the calculation of chiral symmetry, we have used $H(-{\bf k}) = H({\bf k})$. 

\vspace{0.3cm}

\begin{figure}
	\centering
	\includegraphics[width=0.5\linewidth]{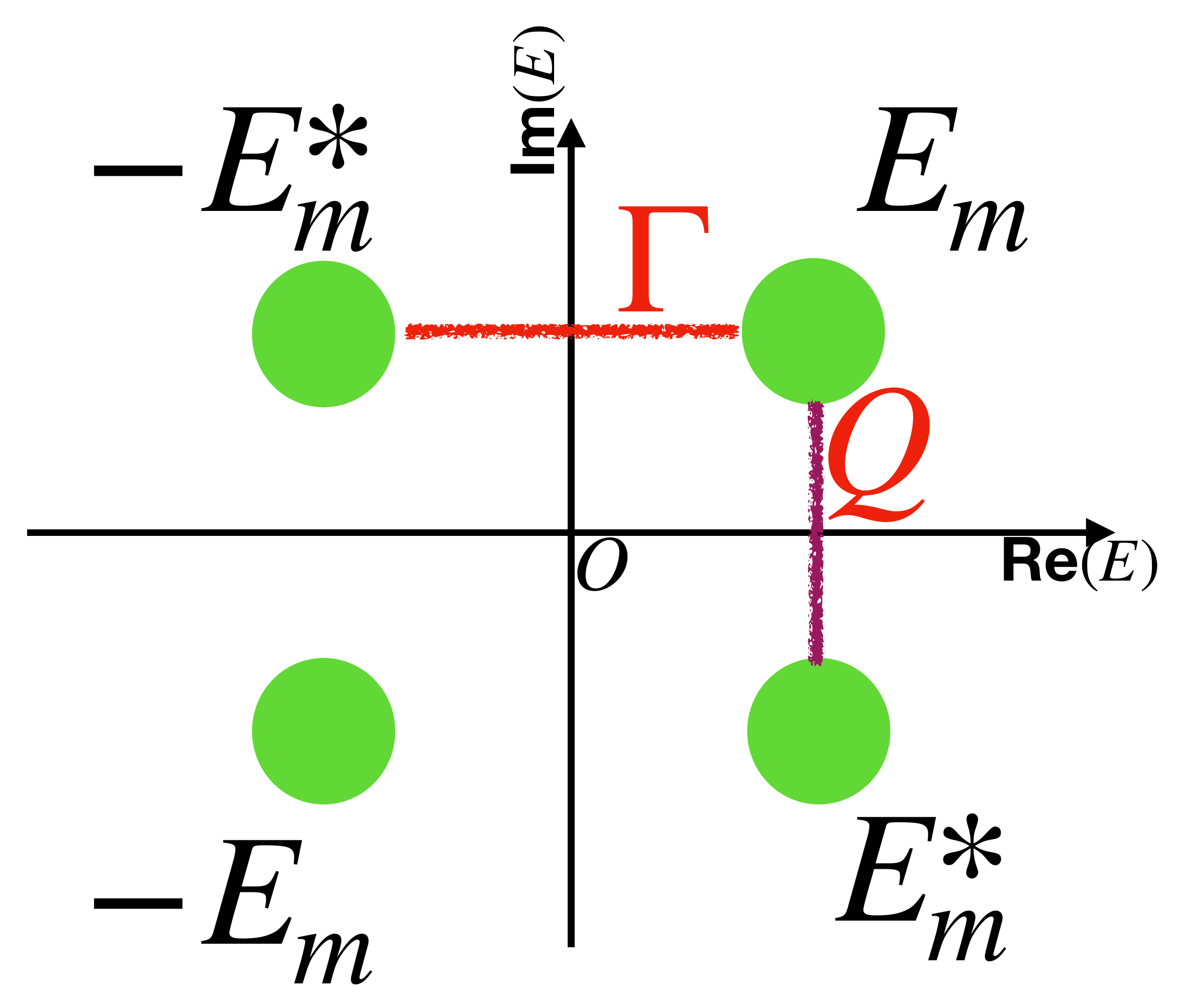}
	\caption{{\bf Effect of different symmetries on the properties of the eigenvalues}. The bi-chiral symmetry ensures $E_n$ and $-E_n^*$ come with pairs;
		while the anti-unitary $Q$ symmetry ensures $E_n$ and $E_n^*$ come with pairs. }
	\label{fig-supp5}
\end{figure}

With this intuition, let's turn to our model with internal $\text{SU}(2)$-symmetry (in the main text we consider the case with $p=1$ and $q = 6$), 
which can be written as the sum of Hermitian part and non-Hermitian part as
\begin{equation}
H = H_h + H_n,
\end{equation}
where 
\begin{equation}
H_h = \tilde{U}^{k_y} \otimes S +  V_s \otimes S^3 + \text{h.c.}+V , \quad H_n =  i \kappa I_{q} \otimes \sigma_z.  
\end{equation}
This model can not be regarded as two independent copies of spinless Haper equation due to the 
non-commutate relation between these two terms. A new chiral operator should be defined as
\begin{equation}
\Gamma \rightarrow \Gamma \otimes {\bf I}_2,
\end{equation}
where in the right-band side $\Gamma$ was defined in Eq. \ref{eq-Gamma}, then we find $\{H_h, \Gamma\} = 0$. However, we noticed that 
$[H_n, \Gamma] = 0$, and $\{H_n, \Gamma\} = 2i\kappa \Gamma \otimes \sigma_z \ne 0$, thus we find that the non-Hermicity can break the conventional chiral symmetry. 
This feature leads to a new symmetry described by the following two commutators,
\begin{equation}
\begin{aligned}
(A,B)_{*} &= AB^\dagger + BA^\dagger, \\
(A,B)^{*} &= A^\dagger B + B^\dagger A. \\
\end{aligned}
\end{equation}
Note that $H = H_h + H_n, H^\dagger = H_h - H_n$. We find 
\begin{equation}
\begin{aligned}
(H,\Gamma)_* &= H\Gamma + \Gamma H^\dagger = \{H,\Gamma\} + [H_n, \Gamma] = 0, \\
(H,\Gamma)^* &= H^\dagger\Gamma + \Gamma H = \{H,\Gamma\} - [H_n, \Gamma] = 0. \\
\end{aligned}
\end{equation}
From the bi-orthogonal relation $H \phi_n^R = E_n \phi_n^R, H^\dagger \phi_n^L = E_n^* \phi_n^L$, then we have,
\begin{equation}
(H,\Gamma)_* \phi_n^L = 0 \to H (\Gamma \phi_n^L) = -E_n^* (\Gamma \phi_n^L).
\end{equation}

If $\Gamma \phi_n^L = \phi_m^R$, we have $E_m = -E_n^*$. In the same spirit ,
\begin{equation}
(H,\Gamma)^*\phi_n^R = 0 \to H^\dagger (\Gamma \phi_n^R) = -E_n (\Gamma \phi_n^R).
\end{equation}
If $\Gamma \phi_n^R = \phi_m^L$ we have $E_m^* = -E_n$. We schematically show the role of this new symmetry in Fig. \ref{fig-supp5}. This symmetry will recover to 
the conventional chiral symmetry in the Hermitian limit with $H^\dagger \rightarrow H$, thus we refer it to bi-chiral symmetry. 

\vspace{0.4cm}
We noticed that the model ($q = 6$) has an extra symmetry in the MBZ defined as
\begin{equation}
Q = {\bf I} \otimes \sigma_x, \quad Q^2 =1.
\end{equation}
This new anti-unitary symmetry (notice that the momentum ${\bf k}$ is unchanged, thus it can not be regarded as the conventional 
time-reversal operator) realizes
\begin{equation}
Q H^* Q = H,
\end{equation}
which guarantees that
\begin{equation}
\bra{\phi^L_m} H^* Q - QH\ket{\phi^R_n}=0, \quad (E_m^* - E_n)\bra{\Phi^L_m}Q\ket{\Phi^R_n} = 0.
\end{equation}
In our model, $\bra{\Phi^L_m}Q\ket{\Phi^R_n} \ne 0$ and $\bra{\Phi^L_n}Q\ket{\Phi^R_n} = 0$, thus we should have $E_m^* = E_n$ if $(m\ne n)$, which is also schematically shown in Fig. \ref{fig-supp5}. 

\begin{figure}[h]
	\centering
	\includegraphics[width=0.9\linewidth]{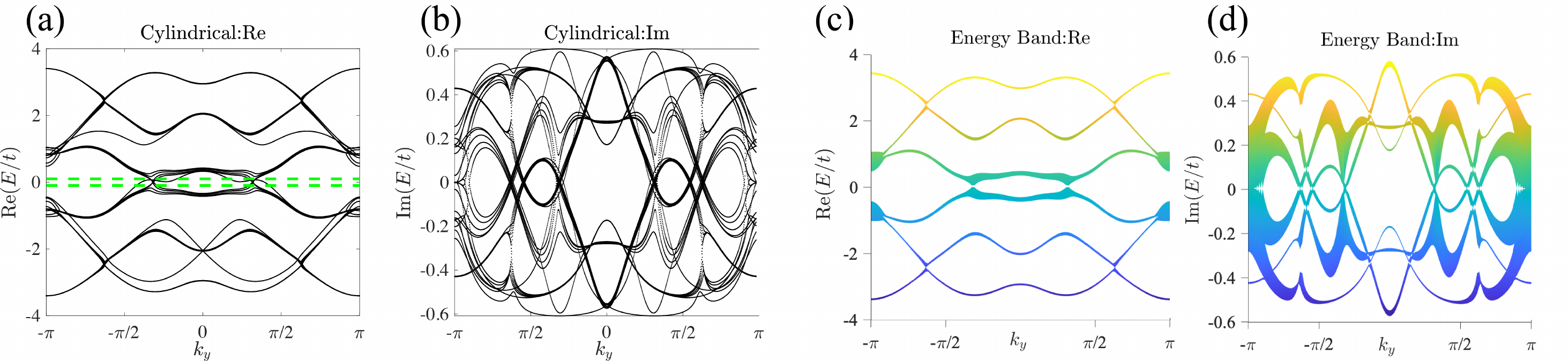}
	\caption{{\bf The details of the energy spectra with $\kappa/t = 0.65$ and $\lambda/t = 0.85$}. (a) and (b) are the real and imaginary eigenvalues in the cylindrical geometry; while 
		(c) and (d) show the corresponding bulk bands.}
	\label{fig-supp6}
\end{figure}

\vspace{0.2cm}
\begin{figure}[h]
	\centering
	\includegraphics[width=0.9\linewidth]{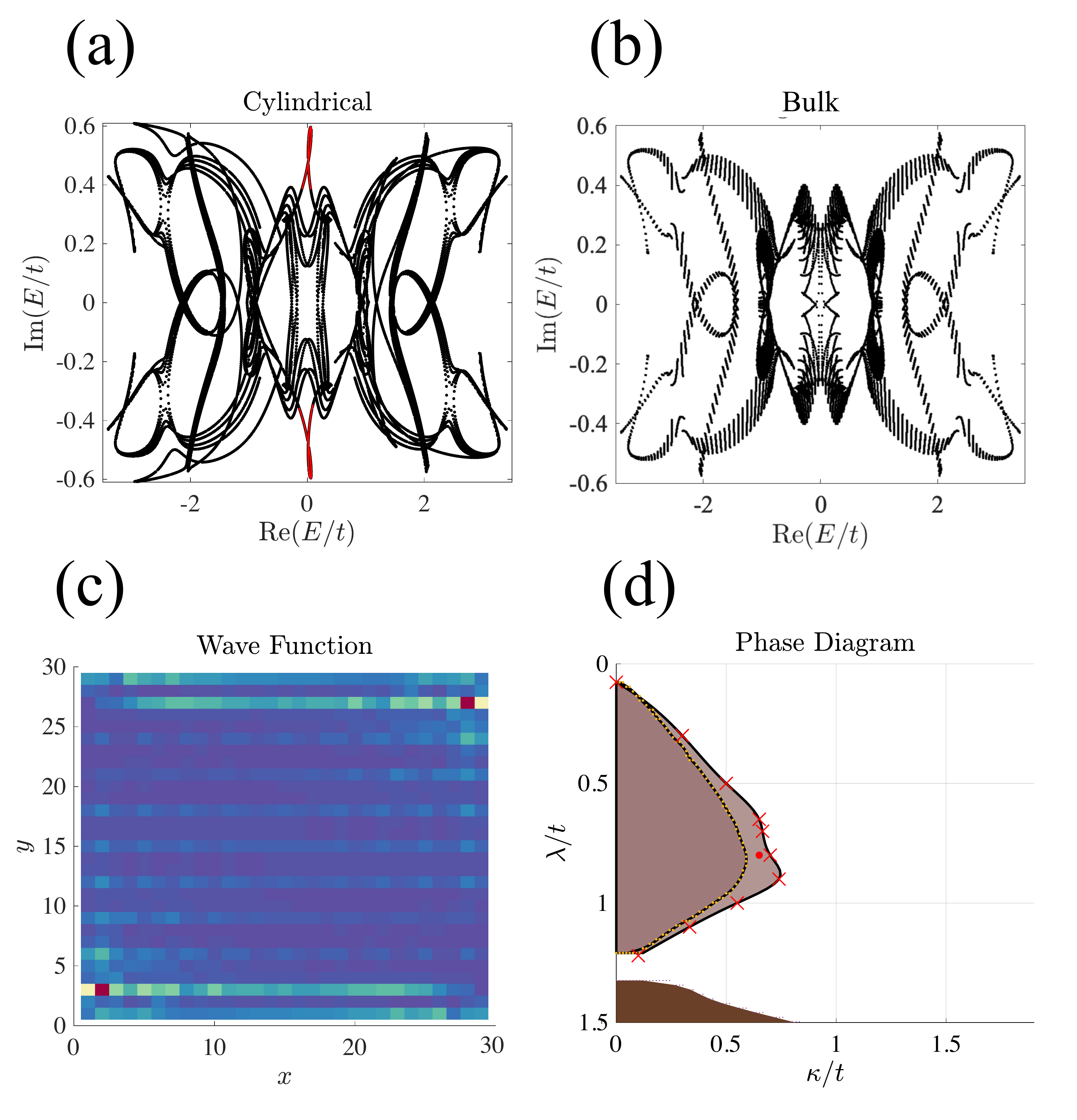}
	\caption{{\bf The details of the energy spectra with $\kappa/t = 0.65$ and $\lambda/t = 0.85$ in the gapless QSH regime}. 
		(a) and (b) The real vs imaginary eigenvalues in cylindrical geometry and the infinite size system, respectively. The robust edge mode is indicated by red lines.
		(c)The wave function of the edge modes in the gapless QSH regime. (d) The phase diagram, where the parameter we choose in (a) - (c) is indicated by a red point.}
	\label{fig-supp7}
\end{figure}

\section{F. Edge modes in the gapless QSH phase regime}

In Fig. \ref{fig-supp6} and Fig. \ref{fig-supp7} we plot the eigenvalues in the above two geometries with non-Hermitian interaction with parameters 
$\kappa/t = 0.65$ and $\lambda/t = 0.85$ (see the red point in 
Fig. \ref{fig-supp6} (d)). In the bulk spectra, the gap is closed and it will not reopen again. Here the closed spectra means that in the complex plane the eigenvalues of these two blocks have some overlap in the complex plane. However, in this case, robust edge modes can still be survived, by circumventing the closed gap in the complex plane. Due to the complex spectra, the edge state is not degenerate 
with the bulk state, as their eigenvalues have different imaginary parts, while their real parts may be the same and vise versa. For this reason, the resonant coupling between the edge modes
and the bulk bands are forbidden and the edge modes can be survived. From the perturbation perspective, the disorder and parameter induced variation can not significantly induce the 
mixing between edge modes and bulk bands, thus the edge modes are robust. 

\vspace{0.4cm}

We need to point out that the bulk bands satisfy the bi-chiral symmetry, time-reversal symmetry and $Q$ symmetry defined above, thus the spectra is always symmetric about real axis and
imaginary axis. However, the edge modes break the $Q$ symmetry while still respects the bi-chiral symmetry, thus the edge modes will exhibit some non-symmetric properties about the 
imaginary axis, as shown in Fig. \ref{fig-supp7} (a). This feature enables us to identify the edge modes and separate them from the bulk bands. 

\section{G. Flying Butterfly Effect}

\begin{figure}[h]
	\centering
	\includegraphics[width=0.9\linewidth]{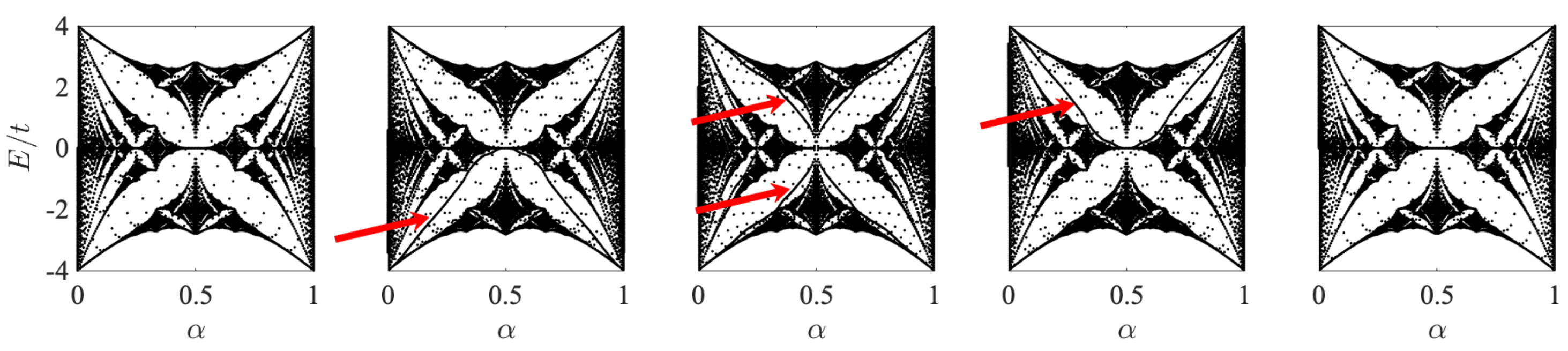}
	\caption{{\bf The details of the Hofstadter butterfly in the spinless model}. Eigenvalues in the cylindrical geometry when $\gamma = 0$ and $k_y$ from $0$ to $2\pi$. In this plot, $k_y$ 
		plays the role of threaded Abelian flux.}
	\label{fig-supp8}
\end{figure}

We will provide some additional details on the flying butterfly effect discussed in the main text. According to the Laughlin's argument, when changing the flux threaded in the cylindrical geometry, 
the effect is that it only move the edge states with finite velocity from one edge to the other edge while the main band structure remains stable. This feature is shown in Fig.~\ref{fig-supp8}, 
which is realized in our simulation by setting $\gamma = 0$. We show that the energy of the edge modes can be changed by the flux represented by $k_y$; however, the background bulk bands with 
fractal properties are unchanged. 

\begin{figure}[h]
	\centering
	\includegraphics[width=0.9\linewidth]{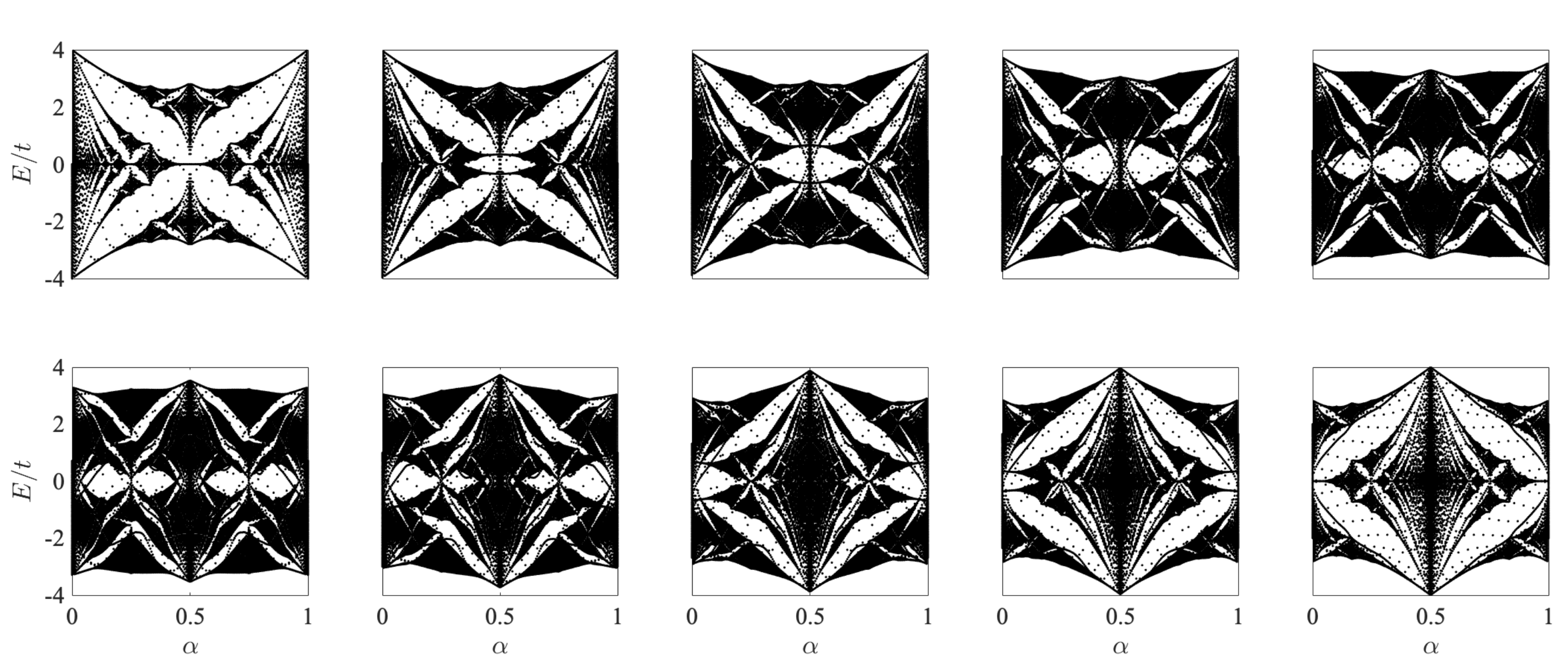}
	\caption{{\bf The details of the Hofstadter butterfly with non-Abelian gauge potential}. Eigenvalues in the cylindrical geometry when $\gamma = 1/4$ (for maximum coupling) and $k_y$ from $0$ to $2\pi$. 
		The spectra is dramatically changed due to the coupling between the two Haper equations by $\gamma$. }
	\label{fig-supp9}
\end{figure}

This picture will be totally changed in the non-Abelian gauge potential due to the direct coupling between the two copies of Abelian Hofstadter butterfly. 
It will give rise to some complex Hofstadter structures by varying the flux $k_y$, as shown in Fig. \ref{fig-supp9}. 
The coupling of these two components have two distinct consequences: (I) The bulk spectra will depend strongly on the coupling strength and the 
flux, thus we find that during the variation of $k_y$, the butterfly also changes dramatically; (II) The coupling can close the band gap, thus the edge modes, 
if existed, maybe emerged in the bulk bands. In this case, the coupling between the edge modes and the extended bulk modes may destroy these edge modes. 
In the main text, we introduce the staggered potential $V_\text{stag}$ to reopen the band gap so as to realize the QSH state. 

% \bibliography{reference.bib}

\begin{thebibliography}{58}%
\makeatletter
\providecommand \@ifxundefined [1]{%
 \@ifx{#1\undefined}
}%
\providecommand \@ifnum [1]{%
 \ifnum #1\expandafter \@firstoftwo
 \else \expandafter \@secondoftwo
 \fi
}%
\providecommand \@ifx [1]{%
 \ifx #1\expandafter \@firstoftwo
 \else \expandafter \@secondoftwo
 \fi
}%
\providecommand \natexlab [1]{#1}%
\providecommand \enquote  [1]{``#1''}%
\providecommand \bibnamefont  [1]{#1}%
\providecommand \bibfnamefont [1]{#1}%
\providecommand \citenamefont [1]{#1}%
\providecommand \href@noop [0]{\@secondoftwo}%
\providecommand \href [0]{\begingroup \@sanitize@url \@href}%
\providecommand \@href[1]{\@@startlink{#1}\@@href}%
\providecommand \@@href[1]{\endgroup#1\@@endlink}%
\providecommand \@sanitize@url [0]{\catcode `\\12\catcode `\$12\catcode
  `\&12\catcode `\#12\catcode `\^12\catcode `\_12\catcode `\%12\relax}%
\providecommand \@@startlink[1]{}%
\providecommand \@@endlink[0]{}%
\providecommand \url  [0]{\begingroup\@sanitize@url \@url }%
\providecommand \@url [1]{\endgroup\@href {#1}{\urlprefix }}%
\providecommand \urlprefix  [0]{URL }%
\providecommand \Eprint [0]{\href }%
\providecommand \doibase [0]{http://dx.doi.org/}%
\providecommand \selectlanguage [0]{\@gobble}%
\providecommand \bibinfo  [0]{\@secondoftwo}%
\providecommand \bibfield  [0]{\@secondoftwo}%
\providecommand \translation [1]{[#1]}%
\providecommand \BibitemOpen [0]{}%
\providecommand \bibitemStop [0]{}%
\providecommand \bibitemNoStop [0]{.\EOS\space}%
\providecommand \EOS [0]{\spacefactor3000\relax}%
\providecommand \BibitemShut  [1]{\csname bibitem#1\endcsname}%
\let\auto@bib@innerbib\@empty
%</preamble>
\bibitem [{\citenamefont {Klitzing}\ \emph {et~al.}(1980)\citenamefont
  {Klitzing}, \citenamefont {Dorda},\ and\ \citenamefont
  {Pepper}}]{klitzing1980new}%
  \BibitemOpen
  \bibfield  {author} {\bibinfo {author} {\bibfnamefont {K.~v.}\ \bibnamefont
  {Klitzing}}, \bibinfo {author} {\bibfnamefont {G.}~\bibnamefont {Dorda}}, \
  and\ \bibinfo {author} {\bibfnamefont {M.}~\bibnamefont {Pepper}},\ }\href
  {\doibase 10.1103/PhysRevLett.45.494} {\bibfield  {journal} {\bibinfo
  {journal} {Phys. Rev. Lett.}\ }\textbf {\bibinfo {volume} {45}},\ \bibinfo
  {pages} {494} (\bibinfo {year} {1980})}\BibitemShut {NoStop}%
\bibitem [{\citenamefont {Tsui}\ \emph {et~al.}(1982)\citenamefont {Tsui},
  \citenamefont {Stormer},\ and\ \citenamefont {Gossard}}]{tsui1982two}%
  \BibitemOpen
  \bibfield  {author} {\bibinfo {author} {\bibfnamefont {D.~C.}\ \bibnamefont
  {Tsui}}, \bibinfo {author} {\bibfnamefont {H.~L.}\ \bibnamefont {Stormer}}, \
  and\ \bibinfo {author} {\bibfnamefont {A.~C.}\ \bibnamefont {Gossard}},\
  }\href {\doibase 10.1103/PhysRevLett.48.1559} {\bibfield  {journal} {\bibinfo
   {journal} {Phys. Rev. Lett.}\ }\textbf {\bibinfo {volume} {48}},\ \bibinfo
  {pages} {1559} (\bibinfo {year} {1982})}\BibitemShut {NoStop}%
\bibitem [{\citenamefont {Laughlin}(1983)}]{laughlin1983}%
  \BibitemOpen
  \bibfield  {author} {\bibinfo {author} {\bibfnamefont {R.~B.}\ \bibnamefont
  {Laughlin}},\ }\href {\doibase 10.1103/PhysRevLett.50.1395} {\bibfield
  {journal} {\bibinfo  {journal} {Phys. Rev. Lett.}\ }\textbf {\bibinfo
  {volume} {50}},\ \bibinfo {pages} {1395} (\bibinfo {year}
  {1983})}\BibitemShut {NoStop}%
\bibitem [{\citenamefont {Ando}(2013)}]{Ando2013topological}%
  \BibitemOpen
  \bibfield  {author} {\bibinfo {author} {\bibfnamefont {Y.}~\bibnamefont
  {Ando}},\ }\href {\doibase 10.7566/JPSJ.82.102001} {\bibfield  {journal}
  {\bibinfo  {journal} {J. Phys. Soc. Jpn}\ }\textbf {\bibinfo {volume} {82}},\
  \bibinfo {pages} {102001} (\bibinfo {year} {2013})}\BibitemShut {NoStop}%
\bibitem [{\citenamefont {Ando}\ and\ \citenamefont
  {Fu}(2015)}]{ando2015topological}%
  \BibitemOpen
  \bibfield  {author} {\bibinfo {author} {\bibfnamefont {Y.}~\bibnamefont
  {Ando}}\ and\ \bibinfo {author} {\bibfnamefont {L.}~\bibnamefont {Fu}},\
  }\href {\doibase 10.1146/annurev-conmatphys-031214-014501} {\bibfield
  {journal} {\bibinfo  {journal} {Annu. Rev. Conden. Ma. Phys.}\ }\textbf
  {\bibinfo {volume} {6}},\ \bibinfo {pages} {361} (\bibinfo {year}
  {2015})}\BibitemShut {NoStop}%
\bibitem [{\citenamefont {Hasan}\ and\ \citenamefont
  {Kane}(2010)}]{hasan2010colloquium}%
  \BibitemOpen
  \bibfield  {author} {\bibinfo {author} {\bibfnamefont {M.~Z.}\ \bibnamefont
  {Hasan}}\ and\ \bibinfo {author} {\bibfnamefont {C.~L.}\ \bibnamefont
  {Kane}},\ }\href {\doibase 10.1103/RevModPhys.82.3045} {\bibfield  {journal}
  {\bibinfo  {journal} {Rev. Mod. Phys.}\ }\textbf {\bibinfo {volume} {82}},\
  \bibinfo {pages} {3045} (\bibinfo {year} {2010})}\BibitemShut {NoStop}%
\bibitem [{\citenamefont {Murakami}\ \emph {et~al.}(2003)\citenamefont
  {Murakami}, \citenamefont {Nagaosa},\ and\ \citenamefont
  {Zhang}}]{murakami2003dissipationless}%
  \BibitemOpen
  \bibfield  {author} {\bibinfo {author} {\bibfnamefont {S.}~\bibnamefont
  {Murakami}}, \bibinfo {author} {\bibfnamefont {N.}~\bibnamefont {Nagaosa}}, \
  and\ \bibinfo {author} {\bibfnamefont {S.-C.}\ \bibnamefont {Zhang}},\ }\href
  {\doibase 10.1126/science.1087128} {\bibfield  {journal} {\bibinfo  {journal}
  {Science}\ }\textbf {\bibinfo {volume} {301}},\ \bibinfo {pages} {1348}
  (\bibinfo {year} {2003})}\BibitemShut {NoStop}%
\bibitem [{\citenamefont {Koralek}\ \emph {et~al.}(2009)\citenamefont
  {Koralek}, \citenamefont {Weber}, \citenamefont {Orenstein}, \citenamefont
  {Bernevig}, \citenamefont {Zhang}, \citenamefont {Mack},\ and\ \citenamefont
  {Awschalom}}]{koralek2009emergence}%
  \BibitemOpen
  \bibfield  {author} {\bibinfo {author} {\bibfnamefont {J.~D.}\ \bibnamefont
  {Koralek}}, \bibinfo {author} {\bibfnamefont {C.~P.}\ \bibnamefont {Weber}},
  \bibinfo {author} {\bibfnamefont {J.}~\bibnamefont {Orenstein}}, \bibinfo
  {author} {\bibfnamefont {B.~A.}\ \bibnamefont {Bernevig}}, \bibinfo {author}
  {\bibfnamefont {S.-C.}\ \bibnamefont {Zhang}}, \bibinfo {author}
  {\bibfnamefont {S.}~\bibnamefont {Mack}}, \ and\ \bibinfo {author}
  {\bibfnamefont {D.}~\bibnamefont {Awschalom}},\ }\href {\doibase
  10.1038/nature07871} {\bibfield  {journal} {\bibinfo  {journal} {Nature}\
  }\textbf {\bibinfo {volume} {458}},\ \bibinfo {pages} {610} (\bibinfo {year}
  {2009})}\BibitemShut {NoStop}%
\bibitem [{\citenamefont {Yuan}\ \emph {et~al.}(2014)\citenamefont {Yuan},
  \citenamefont {Wang}, \citenamefont {Lian}, \citenamefont {Zhang},
  \citenamefont {Fang}, \citenamefont {Shen}, \citenamefont {Xu}, \citenamefont
  {Xu}, \citenamefont {Zhang}, \citenamefont {Hwang} \emph
  {et~al.}}]{yuan2014generation}%
  \BibitemOpen
  \bibfield  {author} {\bibinfo {author} {\bibfnamefont {H.}~\bibnamefont
  {Yuan}}, \bibinfo {author} {\bibfnamefont {X.}~\bibnamefont {Wang}}, \bibinfo
  {author} {\bibfnamefont {B.}~\bibnamefont {Lian}}, \bibinfo {author}
  {\bibfnamefont {H.}~\bibnamefont {Zhang}}, \bibinfo {author} {\bibfnamefont
  {X.}~\bibnamefont {Fang}}, \bibinfo {author} {\bibfnamefont {B.}~\bibnamefont
  {Shen}}, \bibinfo {author} {\bibfnamefont {G.}~\bibnamefont {Xu}}, \bibinfo
  {author} {\bibfnamefont {Y.}~\bibnamefont {Xu}}, \bibinfo {author}
  {\bibfnamefont {S.-C.}\ \bibnamefont {Zhang}}, \bibinfo {author}
  {\bibfnamefont {H.~Y.}\ \bibnamefont {Hwang}},  \emph {et~al.},\ }\href
  {\doibase 10.1038/nnano.2014.183} {\bibfield  {journal} {\bibinfo  {journal}
  {Nat. nanotechnology}\ }\textbf {\bibinfo {volume} {9}},\ \bibinfo {pages}
  {851} (\bibinfo {year} {2014})}\BibitemShut {NoStop}%
\bibitem [{\citenamefont {Hsieh}\ \emph {et~al.}(2009)\citenamefont {Hsieh},
  \citenamefont {Xia}, \citenamefont {Wray}, \citenamefont {Qian},
  \citenamefont {Pal}, \citenamefont {Dil}, \citenamefont {Osterwalder},
  \citenamefont {Meier}, \citenamefont {Bihlmayer}, \citenamefont {Kane},
  \citenamefont {Hor}, \citenamefont {Cava},\ and\ \citenamefont
  {Hasan}}]{hsieh2009observation}%
  \BibitemOpen
  \bibfield  {author} {\bibinfo {author} {\bibfnamefont {D.}~\bibnamefont
  {Hsieh}}, \bibinfo {author} {\bibfnamefont {Y.}~\bibnamefont {Xia}}, \bibinfo
  {author} {\bibfnamefont {L.}~\bibnamefont {Wray}}, \bibinfo {author}
  {\bibfnamefont {D.}~\bibnamefont {Qian}}, \bibinfo {author} {\bibfnamefont
  {A.}~\bibnamefont {Pal}}, \bibinfo {author} {\bibfnamefont {J.~H.}\
  \bibnamefont {Dil}}, \bibinfo {author} {\bibfnamefont {J.}~\bibnamefont
  {Osterwalder}}, \bibinfo {author} {\bibfnamefont {F.}~\bibnamefont {Meier}},
  \bibinfo {author} {\bibfnamefont {G.}~\bibnamefont {Bihlmayer}}, \bibinfo
  {author} {\bibfnamefont {C.~L.}\ \bibnamefont {Kane}}, \bibinfo {author}
  {\bibfnamefont {Y.~S.}\ \bibnamefont {Hor}}, \bibinfo {author} {\bibfnamefont
  {R.~J.}\ \bibnamefont {Cava}}, \ and\ \bibinfo {author} {\bibfnamefont
  {M.~Z.}\ \bibnamefont {Hasan}},\ }\href {\doibase 10.1126/science.1167733}
  {\bibfield  {journal} {\bibinfo  {journal} {Science}\ }\textbf {\bibinfo
  {volume} {323}},\ \bibinfo {pages} {919} (\bibinfo {year}
  {2009})}\BibitemShut {NoStop}%
\bibitem [{\citenamefont {Sau}\ \emph {et~al.}(2010)\citenamefont {Sau},
  \citenamefont {Lutchyn}, \citenamefont {Tewari},\ and\ \citenamefont
  {Das~Sarma}}]{sau2010generic}%
  \BibitemOpen
  \bibfield  {author} {\bibinfo {author} {\bibfnamefont {J.~D.}\ \bibnamefont
  {Sau}}, \bibinfo {author} {\bibfnamefont {R.~M.}\ \bibnamefont {Lutchyn}},
  \bibinfo {author} {\bibfnamefont {S.}~\bibnamefont {Tewari}}, \ and\ \bibinfo
  {author} {\bibfnamefont {S.}~\bibnamefont {Das~Sarma}},\ }\href {\doibase
  10.1103/PhysRevLett.104.040502} {\bibfield  {journal} {\bibinfo  {journal}
  {Phys. Rev. Lett.}\ }\textbf {\bibinfo {volume} {104}},\ \bibinfo {pages}
  {040502} (\bibinfo {year} {2010})}\BibitemShut {NoStop}%
\bibitem [{\citenamefont {Georgescu}\ \emph {et~al.}(2014)\citenamefont
  {Georgescu}, \citenamefont {Ashhab},\ and\ \citenamefont
  {Nori}}]{Georgescu2014}%
  \BibitemOpen
  \bibfield  {author} {\bibinfo {author} {\bibfnamefont {I.~M.}\ \bibnamefont
  {Georgescu}}, \bibinfo {author} {\bibfnamefont {S.}~\bibnamefont {Ashhab}}, \
  and\ \bibinfo {author} {\bibfnamefont {F.}~\bibnamefont {Nori}},\ }\href
  {\doibase 10.1103/RevModPhys.86.153} {\bibfield  {journal} {\bibinfo
  {journal} {Rev. Mod. Phys.}\ }\textbf {\bibinfo {volume} {86}},\ \bibinfo
  {pages} {153} (\bibinfo {year} {2014})}\BibitemShut {NoStop}%
\bibitem [{\citenamefont {Sala}\ \emph {et~al.}(2015)\citenamefont {Sala},
  \citenamefont {Solnyshkov}, \citenamefont {Carusotto}, \citenamefont
  {Jacqmin}, \citenamefont {Lema\^{\i}tre}, \citenamefont
  {Ter\ifmmode~\mbox{\c{c}}\else \c{c}\fi{}as}, \citenamefont {Nalitov},
  \citenamefont {Abbarchi}, \citenamefont {Galopin}, \citenamefont {Sagnes},
  \citenamefont {Bloch}, \citenamefont {Malpuech},\ and\ \citenamefont
  {Amo}}]{sala2015spin}%
  \BibitemOpen
  \bibfield  {author} {\bibinfo {author} {\bibfnamefont {V.~G.}\ \bibnamefont
  {Sala}}, \bibinfo {author} {\bibfnamefont {D.~D.}\ \bibnamefont
  {Solnyshkov}}, \bibinfo {author} {\bibfnamefont {I.}~\bibnamefont
  {Carusotto}}, \bibinfo {author} {\bibfnamefont {T.}~\bibnamefont {Jacqmin}},
  \bibinfo {author} {\bibfnamefont {A.}~\bibnamefont {Lema\^{\i}tre}}, \bibinfo
  {author} {\bibfnamefont {H.}~\bibnamefont {Ter\ifmmode~\mbox{\c{c}}\else
  \c{c}\fi{}as}}, \bibinfo {author} {\bibfnamefont {A.}~\bibnamefont
  {Nalitov}}, \bibinfo {author} {\bibfnamefont {M.}~\bibnamefont {Abbarchi}},
  \bibinfo {author} {\bibfnamefont {E.}~\bibnamefont {Galopin}}, \bibinfo
  {author} {\bibfnamefont {I.}~\bibnamefont {Sagnes}}, \bibinfo {author}
  {\bibfnamefont {J.}~\bibnamefont {Bloch}}, \bibinfo {author} {\bibfnamefont
  {G.}~\bibnamefont {Malpuech}}, \ and\ \bibinfo {author} {\bibfnamefont
  {A.}~\bibnamefont {Amo}},\ }\href {\doibase 10.1103/PhysRevX.5.011034}
  {\bibfield  {journal} {\bibinfo  {journal} {Phys. Rev. X}\ }\textbf {\bibinfo
  {volume} {5}},\ \bibinfo {pages} {011034} (\bibinfo {year}
  {2015})}\BibitemShut {NoStop}%
\bibitem [{\citenamefont {Lu}\ \emph {et~al.}(2014)\citenamefont {Lu},
  \citenamefont {Joannopoulos},\ and\ \citenamefont
  {Solja{\v{c}}i{\'c}}}]{lu2014topological}%
  \BibitemOpen
  \bibfield  {author} {\bibinfo {author} {\bibfnamefont {L.}~\bibnamefont
  {Lu}}, \bibinfo {author} {\bibfnamefont {J.~D.}\ \bibnamefont
  {Joannopoulos}}, \ and\ \bibinfo {author} {\bibfnamefont {M.}~\bibnamefont
  {Solja{\v{c}}i{\'c}}},\ }\href {\doibase 10.1038/nphoton.2014.248} {\bibfield
   {journal} {\bibinfo  {journal} {Nat. Photonics}\ }\textbf {\bibinfo {volume}
  {8}},\ \bibinfo {pages} {821} (\bibinfo {year} {2014})}\BibitemShut {NoStop}%
\bibitem [{\citenamefont {Ozawa}\ \emph {et~al.}(2018)\citenamefont {Ozawa},
  \citenamefont {Price}, \citenamefont {Amo}, \citenamefont {Goldman},
  \citenamefont {Hafezi}, \citenamefont {Lu}, \citenamefont {Rechtsman},
  \citenamefont {Schuster}, \citenamefont {Simon}, \citenamefont {Zilberberg}
  \emph {et~al.}}]{ozawa2018topological}%
  \BibitemOpen
  \bibfield  {author} {\bibinfo {author} {\bibfnamefont {T.}~\bibnamefont
  {Ozawa}}, \bibinfo {author} {\bibfnamefont {H.~M.}\ \bibnamefont {Price}},
  \bibinfo {author} {\bibfnamefont {A.}~\bibnamefont {Amo}}, \bibinfo {author}
  {\bibfnamefont {N.}~\bibnamefont {Goldman}}, \bibinfo {author} {\bibfnamefont
  {M.}~\bibnamefont {Hafezi}}, \bibinfo {author} {\bibfnamefont
  {L.}~\bibnamefont {Lu}}, \bibinfo {author} {\bibfnamefont {M.}~\bibnamefont
  {Rechtsman}}, \bibinfo {author} {\bibfnamefont {D.}~\bibnamefont {Schuster}},
  \bibinfo {author} {\bibfnamefont {J.}~\bibnamefont {Simon}}, \bibinfo
  {author} {\bibfnamefont {O.}~\bibnamefont {Zilberberg}},  \emph {et~al.},\
  }\href {https://arxiv.org/abs/1802.04173} {\bibfield  {journal} {\bibinfo
  {journal} {arXiv:1802.04173}\ } (\bibinfo {year} {2018})}\BibitemShut
  {NoStop}%
\bibitem [{\citenamefont {Khanikaev}\ \emph {et~al.}(2013)\citenamefont
  {Khanikaev}, \citenamefont {Mousavi}, \citenamefont {Tse}, \citenamefont
  {Kargarian}, \citenamefont {MacDonald},\ and\ \citenamefont
  {Shvets}}]{khanikaev2013photonic}%
  \BibitemOpen
  \bibfield  {author} {\bibinfo {author} {\bibfnamefont {A.~B.}\ \bibnamefont
  {Khanikaev}}, \bibinfo {author} {\bibfnamefont {S.~H.}\ \bibnamefont
  {Mousavi}}, \bibinfo {author} {\bibfnamefont {W.-K.}\ \bibnamefont {Tse}},
  \bibinfo {author} {\bibfnamefont {M.}~\bibnamefont {Kargarian}}, \bibinfo
  {author} {\bibfnamefont {A.~H.}\ \bibnamefont {MacDonald}}, \ and\ \bibinfo
  {author} {\bibfnamefont {G.}~\bibnamefont {Shvets}},\ }\href
  {https://doi.org/10.1038/nmat3520} {\bibfield  {journal} {\bibinfo  {journal}
  {Nat. Mat.}\ }\textbf {\bibinfo {volume} {12}},\ \bibinfo {pages} {233}
  (\bibinfo {year} {2013})}\BibitemShut {NoStop}%
\bibitem [{\citenamefont {Cheng}\ \emph {et~al.}(2016)\citenamefont {Cheng},
  \citenamefont {Jouvaud}, \citenamefont {Ni}, \citenamefont {Mousavi},
  \citenamefont {Genack},\ and\ \citenamefont {Khanikaev}}]{cheng2016robust}%
  \BibitemOpen
  \bibfield  {author} {\bibinfo {author} {\bibfnamefont {X.}~\bibnamefont
  {Cheng}}, \bibinfo {author} {\bibfnamefont {C.}~\bibnamefont {Jouvaud}},
  \bibinfo {author} {\bibfnamefont {X.}~\bibnamefont {Ni}}, \bibinfo {author}
  {\bibfnamefont {S.~H.}\ \bibnamefont {Mousavi}}, \bibinfo {author}
  {\bibfnamefont {A.~Z.}\ \bibnamefont {Genack}}, \ and\ \bibinfo {author}
  {\bibfnamefont {A.~B.}\ \bibnamefont {Khanikaev}},\ }\href
  {https://doi.org/10.1038/nmat4573} {\bibfield  {journal} {\bibinfo  {journal}
  {Nat. Mat.}\ }\textbf {\bibinfo {volume} {15}},\ \bibinfo {pages} {542}
  (\bibinfo {year} {2016})}\BibitemShut {NoStop}%
\bibitem [{\citenamefont {Zhang}\ \emph {et~al.}(2013)\citenamefont {Zhang},
  \citenamefont {Ren}, \citenamefont {Wang},\ and\ \citenamefont
  {Li}}]{zhang2013topological}%
  \BibitemOpen
  \bibfield  {author} {\bibinfo {author} {\bibfnamefont {L.}~\bibnamefont
  {Zhang}}, \bibinfo {author} {\bibfnamefont {J.}~\bibnamefont {Ren}}, \bibinfo
  {author} {\bibfnamefont {J.-S.}\ \bibnamefont {Wang}}, \ and\ \bibinfo
  {author} {\bibfnamefont {B.}~\bibnamefont {Li}},\ }\href {\doibase
  10.1103/PhysRevB.87.144101} {\bibfield  {journal} {\bibinfo  {journal} {Phys.
  Rev. B}\ }\textbf {\bibinfo {volume} {87}},\ \bibinfo {pages} {144101}
  (\bibinfo {year} {2013})}\BibitemShut {NoStop}%
\bibitem [{\citenamefont {Mochizuki}\ \emph {et~al.}(2014)\citenamefont
  {Mochizuki}, \citenamefont {Yu}, \citenamefont {Seki}, \citenamefont
  {Kanazawa}, \citenamefont {Koshibae}, \citenamefont {Zang}, \citenamefont
  {Mostovoy}, \citenamefont {Tokura},\ and\ \citenamefont
  {Nagaosa}}]{mochizuki2014thermally}%
  \BibitemOpen
  \bibfield  {author} {\bibinfo {author} {\bibfnamefont {M.}~\bibnamefont
  {Mochizuki}}, \bibinfo {author} {\bibfnamefont {X.}~\bibnamefont {Yu}},
  \bibinfo {author} {\bibfnamefont {S.}~\bibnamefont {Seki}}, \bibinfo {author}
  {\bibfnamefont {N.}~\bibnamefont {Kanazawa}}, \bibinfo {author}
  {\bibfnamefont {W.}~\bibnamefont {Koshibae}}, \bibinfo {author}
  {\bibfnamefont {J.}~\bibnamefont {Zang}}, \bibinfo {author} {\bibfnamefont
  {M.}~\bibnamefont {Mostovoy}}, \bibinfo {author} {\bibfnamefont
  {Y.}~\bibnamefont {Tokura}}, \ and\ \bibinfo {author} {\bibfnamefont
  {N.}~\bibnamefont {Nagaosa}},\ }\href {\doibase 10.1038/nmat3862} {\bibfield
  {journal} {\bibinfo  {journal} {Nat. Mat.}\ }\textbf {\bibinfo {volume}
  {13}},\ \bibinfo {pages} {241} (\bibinfo {year} {2014})}\BibitemShut
  {NoStop}%
\bibitem [{\citenamefont {Aidelsburger}\ \emph {et~al.}(2013)\citenamefont
  {Aidelsburger}, \citenamefont {Atala}, \citenamefont {Lohse}, \citenamefont
  {Barreiro}, \citenamefont {Paredes},\ and\ \citenamefont
  {Bloch}}]{aidelsburger2013realization}%
  \BibitemOpen
  \bibfield  {author} {\bibinfo {author} {\bibfnamefont {M.}~\bibnamefont
  {Aidelsburger}}, \bibinfo {author} {\bibfnamefont {M.}~\bibnamefont {Atala}},
  \bibinfo {author} {\bibfnamefont {M.}~\bibnamefont {Lohse}}, \bibinfo
  {author} {\bibfnamefont {J.~T.}\ \bibnamefont {Barreiro}}, \bibinfo {author}
  {\bibfnamefont {B.}~\bibnamefont {Paredes}}, \ and\ \bibinfo {author}
  {\bibfnamefont {I.}~\bibnamefont {Bloch}},\ }\href {\doibase
  10.1103/PhysRevLett.111.185301} {\bibfield  {journal} {\bibinfo  {journal}
  {Phys. Rev. Lett.}\ }\textbf {\bibinfo {volume} {111}},\ \bibinfo {pages}
  {185301} (\bibinfo {year} {2013})}\BibitemShut {NoStop}%
\bibitem [{\citenamefont {Goldman}\ \emph {et~al.}(2009)\citenamefont
  {Goldman}, \citenamefont {Kubasiak}, \citenamefont {Bermudez}, \citenamefont
  {Gaspard}, \citenamefont {Lewenstein},\ and\ \citenamefont
  {Martin-Delgado}}]{Goldman2009}%
  \BibitemOpen
  \bibfield  {author} {\bibinfo {author} {\bibfnamefont {N.}~\bibnamefont
  {Goldman}}, \bibinfo {author} {\bibfnamefont {A.}~\bibnamefont {Kubasiak}},
  \bibinfo {author} {\bibfnamefont {A.}~\bibnamefont {Bermudez}}, \bibinfo
  {author} {\bibfnamefont {P.}~\bibnamefont {Gaspard}}, \bibinfo {author}
  {\bibfnamefont {M.}~\bibnamefont {Lewenstein}}, \ and\ \bibinfo {author}
  {\bibfnamefont {M.~A.}\ \bibnamefont {Martin-Delgado}},\ }\href {\doibase
  10.1103/PhysRevLett.103.035301} {\bibfield  {journal} {\bibinfo  {journal}
  {Phys. Rev. Lett.}\ }\textbf {\bibinfo {volume} {103}},\ \bibinfo {pages}
  {035301} (\bibinfo {year} {2009})}\BibitemShut {NoStop}%
\bibitem [{\citenamefont {Aidelsburger}\ \emph {et~al.}(2011)\citenamefont
  {Aidelsburger}, \citenamefont {Atala}, \citenamefont {Nascimb\`ene},
  \citenamefont {Trotzky}, \citenamefont {Chen},\ and\ \citenamefont
  {Bloch}}]{aidelsburger2011experimental}%
  \BibitemOpen
  \bibfield  {author} {\bibinfo {author} {\bibfnamefont {M.}~\bibnamefont
  {Aidelsburger}}, \bibinfo {author} {\bibfnamefont {M.}~\bibnamefont {Atala}},
  \bibinfo {author} {\bibfnamefont {S.}~\bibnamefont {Nascimb\`ene}}, \bibinfo
  {author} {\bibfnamefont {S.}~\bibnamefont {Trotzky}}, \bibinfo {author}
  {\bibfnamefont {Y.-A.}\ \bibnamefont {Chen}}, \ and\ \bibinfo {author}
  {\bibfnamefont {I.}~\bibnamefont {Bloch}},\ }\href {\doibase
  10.1103/PhysRevLett.107.255301} {\bibfield  {journal} {\bibinfo  {journal}
  {Phys. Rev. Lett.}\ }\textbf {\bibinfo {volume} {107}},\ \bibinfo {pages}
  {255301} (\bibinfo {year} {2011})}\BibitemShut {NoStop}%
\bibitem [{\citenamefont {Aidelsburger}\ \emph {et~al.}(2015)\citenamefont
  {Aidelsburger}, \citenamefont {Lohse}, \citenamefont {Schweizer},
  \citenamefont {Atala}, \citenamefont {Barreiro}, \citenamefont {Nascimbene},
  \citenamefont {Cooper}, \citenamefont {Bloch},\ and\ \citenamefont
  {Goldman}}]{aidelsburger2015measuring}%
  \BibitemOpen
  \bibfield  {author} {\bibinfo {author} {\bibfnamefont {M.}~\bibnamefont
  {Aidelsburger}}, \bibinfo {author} {\bibfnamefont {M.}~\bibnamefont {Lohse}},
  \bibinfo {author} {\bibfnamefont {C.}~\bibnamefont {Schweizer}}, \bibinfo
  {author} {\bibfnamefont {M.}~\bibnamefont {Atala}}, \bibinfo {author}
  {\bibfnamefont {J.~T.}\ \bibnamefont {Barreiro}}, \bibinfo {author}
  {\bibfnamefont {S.}~\bibnamefont {Nascimbene}}, \bibinfo {author}
  {\bibfnamefont {N.}~\bibnamefont {Cooper}}, \bibinfo {author} {\bibfnamefont
  {I.}~\bibnamefont {Bloch}}, \ and\ \bibinfo {author} {\bibfnamefont
  {N.}~\bibnamefont {Goldman}},\ }\href {\doibase 10.1038/nphys3171} {\bibfield
   {journal} {\bibinfo  {journal} {Nat. Phys.}\ }\textbf {\bibinfo {volume}
  {11}},\ \bibinfo {pages} {162} (\bibinfo {year} {2015})}\BibitemShut
  {NoStop}%
\bibitem [{\citenamefont {Ningyuan}\ \emph {et~al.}(2015)\citenamefont
  {Ningyuan}, \citenamefont {Owens}, \citenamefont {Sommer}, \citenamefont
  {Schuster},\ and\ \citenamefont {Simon}}]{Ningyuan2015}%
  \BibitemOpen
  \bibfield  {author} {\bibinfo {author} {\bibfnamefont {J.}~\bibnamefont
  {Ningyuan}}, \bibinfo {author} {\bibfnamefont {C.}~\bibnamefont {Owens}},
  \bibinfo {author} {\bibfnamefont {A.}~\bibnamefont {Sommer}}, \bibinfo
  {author} {\bibfnamefont {D.}~\bibnamefont {Schuster}}, \ and\ \bibinfo
  {author} {\bibfnamefont {J.}~\bibnamefont {Simon}},\ }\href {\doibase
  10.1103/PhysRevX.5.021031} {\bibfield  {journal} {\bibinfo  {journal} {Phys.
  Rev. X}\ }\textbf {\bibinfo {volume} {5}},\ \bibinfo {pages} {021031}
  (\bibinfo {year} {2015})}\BibitemShut {NoStop}%
\bibitem [{\citenamefont {Imhof}\ \emph {et~al.}(2018)\citenamefont {Imhof},
  \citenamefont {Berger}, \citenamefont {Bayer}, \citenamefont {Brehm},
  \citenamefont {Molenkamp}, \citenamefont {Kiessling}, \citenamefont
  {Schindler}, \citenamefont {Lee}, \citenamefont {Greiter}, \citenamefont
  {Neupert} \emph {et~al.}}]{imhof2018topolectrical}%
  \BibitemOpen
  \bibfield  {author} {\bibinfo {author} {\bibfnamefont {S.}~\bibnamefont
  {Imhof}}, \bibinfo {author} {\bibfnamefont {C.}~\bibnamefont {Berger}},
  \bibinfo {author} {\bibfnamefont {F.}~\bibnamefont {Bayer}}, \bibinfo
  {author} {\bibfnamefont {J.}~\bibnamefont {Brehm}}, \bibinfo {author}
  {\bibfnamefont {L.~W.}\ \bibnamefont {Molenkamp}}, \bibinfo {author}
  {\bibfnamefont {T.}~\bibnamefont {Kiessling}}, \bibinfo {author}
  {\bibfnamefont {F.}~\bibnamefont {Schindler}}, \bibinfo {author}
  {\bibfnamefont {C.~H.}\ \bibnamefont {Lee}}, \bibinfo {author} {\bibfnamefont
  {M.}~\bibnamefont {Greiter}}, \bibinfo {author} {\bibfnamefont
  {T.}~\bibnamefont {Neupert}},  \emph {et~al.},\ }\href {\doibase
  10.1038/s41567-018-0246-1} {\bibfield  {journal} {\bibinfo  {journal} {Nat.
  Phys.}\ }\textbf {\bibinfo {volume} {14}},\ \bibinfo {pages} {925} (\bibinfo
  {year} {2018})}\BibitemShut {NoStop}%
\bibitem [{\citenamefont {Zhu}\ \emph {et~al.}(2018)\citenamefont {Zhu},
  \citenamefont {Hou}, \citenamefont {Long}, \citenamefont {Chen},\ and\
  \citenamefont {Ren}}]{Zhu2018}%
  \BibitemOpen
  \bibfield  {author} {\bibinfo {author} {\bibfnamefont {W.}~\bibnamefont
  {Zhu}}, \bibinfo {author} {\bibfnamefont {S.}~\bibnamefont {Hou}}, \bibinfo
  {author} {\bibfnamefont {Y.}~\bibnamefont {Long}}, \bibinfo {author}
  {\bibfnamefont {H.}~\bibnamefont {Chen}}, \ and\ \bibinfo {author}
  {\bibfnamefont {J.}~\bibnamefont {Ren}},\ }\href {\doibase
  10.1103/PhysRevB.97.075310} {\bibfield  {journal} {\bibinfo  {journal} {Phys.
  Rev. B}\ }\textbf {\bibinfo {volume} {97}},\ \bibinfo {pages} {075310}
  (\bibinfo {year} {2018})}\BibitemShut {NoStop}%
\bibitem [{\citenamefont {Roushan}\ \emph {et~al.}(2017)\citenamefont
  {Roushan}, \citenamefont {Neill}, \citenamefont {Megrant}, \citenamefont
  {Chen}, \citenamefont {Babbush}, \citenamefont {Barends}, \citenamefont
  {Campbell}, \citenamefont {Chen}, \citenamefont {Chiaro}, \citenamefont
  {Dunsworth}, \citenamefont {Fowler}, \citenamefont {Jeffrey}, \citenamefont
  {Kelly}, \citenamefont {Lucero}, \citenamefont {Mutus}, \citenamefont
  {O'Malley}, \citenamefont {Neeley}, \citenamefont {Quintana}, \citenamefont
  {Sank}, \citenamefont {Vainsencher}, \citenamefont {Wenner}, \citenamefont
  {White}, \citenamefont {Kapit}, \citenamefont {Neven},\ and\ \citenamefont
  {Martinis}}]{Roushan2017a}%
  \BibitemOpen
  \bibfield  {author} {\bibinfo {author} {\bibfnamefont {P.}~\bibnamefont
  {Roushan}}, \bibinfo {author} {\bibfnamefont {C.}~\bibnamefont {Neill}},
  \bibinfo {author} {\bibfnamefont {A.}~\bibnamefont {Megrant}}, \bibinfo
  {author} {\bibfnamefont {Y.}~\bibnamefont {Chen}}, \bibinfo {author}
  {\bibfnamefont {R.}~\bibnamefont {Babbush}}, \bibinfo {author} {\bibfnamefont
  {R.}~\bibnamefont {Barends}}, \bibinfo {author} {\bibfnamefont
  {B.}~\bibnamefont {Campbell}}, \bibinfo {author} {\bibfnamefont
  {Z.}~\bibnamefont {Chen}}, \bibinfo {author} {\bibfnamefont {B.}~\bibnamefont
  {Chiaro}}, \bibinfo {author} {\bibfnamefont {A.}~\bibnamefont {Dunsworth}},
  \bibinfo {author} {\bibfnamefont {A.}~\bibnamefont {Fowler}}, \bibinfo
  {author} {\bibfnamefont {E.}~\bibnamefont {Jeffrey}}, \bibinfo {author}
  {\bibfnamefont {J.}~\bibnamefont {Kelly}}, \bibinfo {author} {\bibfnamefont
  {E.}~\bibnamefont {Lucero}}, \bibinfo {author} {\bibfnamefont
  {J.}~\bibnamefont {Mutus}}, \bibinfo {author} {\bibfnamefont {P.~J.}\
  \bibnamefont {O'Malley}}, \bibinfo {author} {\bibfnamefont {M.}~\bibnamefont
  {Neeley}}, \bibinfo {author} {\bibfnamefont {C.}~\bibnamefont {Quintana}},
  \bibinfo {author} {\bibfnamefont {D.}~\bibnamefont {Sank}}, \bibinfo {author}
  {\bibfnamefont {A.}~\bibnamefont {Vainsencher}}, \bibinfo {author}
  {\bibfnamefont {J.}~\bibnamefont {Wenner}}, \bibinfo {author} {\bibfnamefont
  {T.}~\bibnamefont {White}}, \bibinfo {author} {\bibfnamefont
  {E.}~\bibnamefont {Kapit}}, \bibinfo {author} {\bibfnamefont
  {H.}~\bibnamefont {Neven}}, \ and\ \bibinfo {author} {\bibfnamefont
  {J.}~\bibnamefont {Martinis}},\ }\href {\doibase 10.1038/nphys3930}
  {\bibfield  {journal} {\bibinfo  {journal} {Nat. Phys.}\ }\textbf {\bibinfo
  {volume} {13}},\ \bibinfo {pages} {146} (\bibinfo {year} {2017})}\BibitemShut
  {NoStop}%
\bibitem [{\citenamefont {Lu}\ \emph {et~al.}(2015)\citenamefont {Lu},
  \citenamefont {Wang}, \citenamefont {Ye}, \citenamefont {Ran}, \citenamefont
  {Fu}, \citenamefont {Joannopoulos},\ and\ \citenamefont {Solja{\v
  c}i{\'c}}}]{lu2015experimental}%
  \BibitemOpen
  \bibfield  {author} {\bibinfo {author} {\bibfnamefont {L.}~\bibnamefont
  {Lu}}, \bibinfo {author} {\bibfnamefont {Z.}~\bibnamefont {Wang}}, \bibinfo
  {author} {\bibfnamefont {D.}~\bibnamefont {Ye}}, \bibinfo {author}
  {\bibfnamefont {L.}~\bibnamefont {Ran}}, \bibinfo {author} {\bibfnamefont
  {L.}~\bibnamefont {Fu}}, \bibinfo {author} {\bibfnamefont {J.~D.}\
  \bibnamefont {Joannopoulos}}, \ and\ \bibinfo {author} {\bibfnamefont
  {M.}~\bibnamefont {Solja{\v c}i{\'c}}},\ }\href
  {http://science.sciencemag.org/content/349/6248/622} {\bibfield  {journal}
  {\bibinfo  {journal} {Science}\ }\textbf {\bibinfo {volume} {349}},\ \bibinfo
  {pages} {622} (\bibinfo {year} {2015})}\BibitemShut {NoStop}%
\bibitem [{\citenamefont {Yang}\ \emph {et~al.}(2018)\citenamefont {Yang},
  \citenamefont {Guo}, \citenamefont {Tremain}, \citenamefont {Liu},
  \citenamefont {Barr}, \citenamefont {Yan}, \citenamefont {Gao}, \citenamefont
  {Liu}, \citenamefont {Xiang}, \citenamefont {Chen}, \citenamefont {Fang},
  \citenamefont {Hibbins}, \citenamefont {Lu},\ and\ \citenamefont
  {Zhang}}]{yang2018ideal}%
  \BibitemOpen
  \bibfield  {author} {\bibinfo {author} {\bibfnamefont {B.}~\bibnamefont
  {Yang}}, \bibinfo {author} {\bibfnamefont {Q.}~\bibnamefont {Guo}}, \bibinfo
  {author} {\bibfnamefont {B.}~\bibnamefont {Tremain}}, \bibinfo {author}
  {\bibfnamefont {R.}~\bibnamefont {Liu}}, \bibinfo {author} {\bibfnamefont
  {L.~E.}\ \bibnamefont {Barr}}, \bibinfo {author} {\bibfnamefont
  {Q.}~\bibnamefont {Yan}}, \bibinfo {author} {\bibfnamefont {W.}~\bibnamefont
  {Gao}}, \bibinfo {author} {\bibfnamefont {H.}~\bibnamefont {Liu}}, \bibinfo
  {author} {\bibfnamefont {Y.}~\bibnamefont {Xiang}}, \bibinfo {author}
  {\bibfnamefont {J.}~\bibnamefont {Chen}}, \bibinfo {author} {\bibfnamefont
  {C.}~\bibnamefont {Fang}}, \bibinfo {author} {\bibfnamefont {A.}~\bibnamefont
  {Hibbins}}, \bibinfo {author} {\bibfnamefont {L.}~\bibnamefont {Lu}}, \ and\
  \bibinfo {author} {\bibfnamefont {S.}~\bibnamefont {Zhang}},\ }\href
  {\doibase 10.1126/science.aaq1221} {\bibfield  {journal} {\bibinfo  {journal}
  {Science}\ }\textbf {\bibinfo {volume} {359}},\ \bibinfo {pages} {1013}
  (\bibinfo {year} {2018})}\BibitemShut {NoStop}%
\bibitem [{\citenamefont {St-Jean}\ \emph {et~al.}(2017)\citenamefont
  {St-Jean}, \citenamefont {Goblot}, \citenamefont {Galopin}, \citenamefont
  {Lema{\^\i}tre}, \citenamefont {Ozawa}, \citenamefont {Le~Gratiet},
  \citenamefont {Sagnes}, \citenamefont {Bloch},\ and\ \citenamefont
  {Amo}}]{st2017lasing}%
  \BibitemOpen
  \bibfield  {author} {\bibinfo {author} {\bibfnamefont {P.}~\bibnamefont
  {St-Jean}}, \bibinfo {author} {\bibfnamefont {V.}~\bibnamefont {Goblot}},
  \bibinfo {author} {\bibfnamefont {E.}~\bibnamefont {Galopin}}, \bibinfo
  {author} {\bibfnamefont {A.}~\bibnamefont {Lema{\^\i}tre}}, \bibinfo {author}
  {\bibfnamefont {T.}~\bibnamefont {Ozawa}}, \bibinfo {author} {\bibfnamefont
  {L.}~\bibnamefont {Le~Gratiet}}, \bibinfo {author} {\bibfnamefont
  {I.}~\bibnamefont {Sagnes}}, \bibinfo {author} {\bibfnamefont
  {J.}~\bibnamefont {Bloch}}, \ and\ \bibinfo {author} {\bibfnamefont
  {A.}~\bibnamefont {Amo}},\ }\href {\doibase 10.1038/s41566-017-0006-2}
  {\bibfield  {journal} {\bibinfo  {journal} {Nat. Photonics}\ }\textbf
  {\bibinfo {volume} {11}},\ \bibinfo {pages} {651} (\bibinfo {year}
  {2017})}\BibitemShut {NoStop}%
\bibitem [{\citenamefont {Harari}\ \emph {et~al.}(2018)\citenamefont {Harari},
  \citenamefont {Bandres}, \citenamefont {Lumer}, \citenamefont {Rechtsman},
  \citenamefont {Chong}, \citenamefont {Khajavikhan}, \citenamefont
  {Christodoulides},\ and\ \citenamefont {Segev}}]{harari2018topological}%
  \BibitemOpen
  \bibfield  {author} {\bibinfo {author} {\bibfnamefont {G.}~\bibnamefont
  {Harari}}, \bibinfo {author} {\bibfnamefont {M.~A.}\ \bibnamefont {Bandres}},
  \bibinfo {author} {\bibfnamefont {Y.}~\bibnamefont {Lumer}}, \bibinfo
  {author} {\bibfnamefont {M.~C.}\ \bibnamefont {Rechtsman}}, \bibinfo {author}
  {\bibfnamefont {Y.~D.}\ \bibnamefont {Chong}}, \bibinfo {author}
  {\bibfnamefont {M.}~\bibnamefont {Khajavikhan}}, \bibinfo {author}
  {\bibfnamefont {D.~N.}\ \bibnamefont {Christodoulides}}, \ and\ \bibinfo
  {author} {\bibfnamefont {M.}~\bibnamefont {Segev}},\ }\href
  {http://science.sciencemag.org/content/359/6381/eaar4003} {\bibfield
  {journal} {\bibinfo  {journal} {Science}\ }\textbf {\bibinfo {volume} {359}}
  (\bibinfo {year} {2018})}\BibitemShut {NoStop}%
\bibitem [{\citenamefont {Paik}\ \emph {et~al.}(2011)\citenamefont {Paik},
  \citenamefont {Schuster}, \citenamefont {Bishop}, \citenamefont {Kirchmair},
  \citenamefont {Catelani}, \citenamefont {Sears}, \citenamefont {Johnson},
  \citenamefont {Reagor}, \citenamefont {Frunzio}, \citenamefont {Glazman},
  \citenamefont {Girvin}, \citenamefont {Devoret},\ and\ \citenamefont
  {Schoelkopf}}]{Paik2011}%
  \BibitemOpen
  \bibfield  {author} {\bibinfo {author} {\bibfnamefont {H.}~\bibnamefont
  {Paik}}, \bibinfo {author} {\bibfnamefont {D.~I.}\ \bibnamefont {Schuster}},
  \bibinfo {author} {\bibfnamefont {L.~S.}\ \bibnamefont {Bishop}}, \bibinfo
  {author} {\bibfnamefont {G.}~\bibnamefont {Kirchmair}}, \bibinfo {author}
  {\bibfnamefont {G.}~\bibnamefont {Catelani}}, \bibinfo {author}
  {\bibfnamefont {A.~P.}\ \bibnamefont {Sears}}, \bibinfo {author}
  {\bibfnamefont {B.~R.}\ \bibnamefont {Johnson}}, \bibinfo {author}
  {\bibfnamefont {M.~J.}\ \bibnamefont {Reagor}}, \bibinfo {author}
  {\bibfnamefont {L.}~\bibnamefont {Frunzio}}, \bibinfo {author} {\bibfnamefont
  {L.~I.}\ \bibnamefont {Glazman}}, \bibinfo {author} {\bibfnamefont {S.~M.}\
  \bibnamefont {Girvin}}, \bibinfo {author} {\bibfnamefont {M.~H.}\
  \bibnamefont {Devoret}}, \ and\ \bibinfo {author} {\bibfnamefont {R.~J.}\
  \bibnamefont {Schoelkopf}},\ }\href {\doibase 10.1103/PhysRevLett.107.240501}
  {\bibfield  {journal} {\bibinfo  {journal} {Phys. Rev. Lett.}\ }\textbf
  {\bibinfo {volume} {107}},\ \bibinfo {pages} {240501} (\bibinfo {year}
  {2011})}\BibitemShut {NoStop}%
\bibitem [{\citenamefont {Harrow}\ and\ \citenamefont
  {Montanaro}(2017)}]{harrow2017quantum}%
  \BibitemOpen
  \bibfield  {author} {\bibinfo {author} {\bibfnamefont {A.~W.}\ \bibnamefont
  {Harrow}}\ and\ \bibinfo {author} {\bibfnamefont {A.}~\bibnamefont
  {Montanaro}},\ }\href {\doibase 10.1038/nature23458} {\bibfield  {journal}
  {\bibinfo  {journal} {Nature}\ }\textbf {\bibinfo {volume} {549}},\ \bibinfo
  {pages} {203} (\bibinfo {year} {2017})}\BibitemShut {NoStop}%
\bibitem [{\citenamefont {Song}\ \emph {et~al.}(2017)\citenamefont {Song},
  \citenamefont {Xu}, \citenamefont {Liu}, \citenamefont {Yang}, \citenamefont
  {Zheng}, \citenamefont {Deng}, \citenamefont {Xie}, \citenamefont {Huang},
  \citenamefont {Guo}, \citenamefont {Zhang}, \citenamefont {Zhang},
  \citenamefont {Xu}, \citenamefont {Zheng}, \citenamefont {Zhu}, \citenamefont
  {Wang}, \citenamefont {Chen}, \citenamefont {Lu}, \citenamefont {Han},\ and\
  \citenamefont {Pan}}]{song201710}%
  \BibitemOpen
  \bibfield  {author} {\bibinfo {author} {\bibfnamefont {C.}~\bibnamefont
  {Song}}, \bibinfo {author} {\bibfnamefont {K.}~\bibnamefont {Xu}}, \bibinfo
  {author} {\bibfnamefont {W.}~\bibnamefont {Liu}}, \bibinfo {author}
  {\bibfnamefont {C.-p.}\ \bibnamefont {Yang}}, \bibinfo {author}
  {\bibfnamefont {S.-B.}\ \bibnamefont {Zheng}}, \bibinfo {author}
  {\bibfnamefont {H.}~\bibnamefont {Deng}}, \bibinfo {author} {\bibfnamefont
  {Q.}~\bibnamefont {Xie}}, \bibinfo {author} {\bibfnamefont {K.}~\bibnamefont
  {Huang}}, \bibinfo {author} {\bibfnamefont {Q.}~\bibnamefont {Guo}}, \bibinfo
  {author} {\bibfnamefont {L.}~\bibnamefont {Zhang}}, \bibinfo {author}
  {\bibfnamefont {P.}~\bibnamefont {Zhang}}, \bibinfo {author} {\bibfnamefont
  {D.}~\bibnamefont {Xu}}, \bibinfo {author} {\bibfnamefont {D.}~\bibnamefont
  {Zheng}}, \bibinfo {author} {\bibfnamefont {X.}~\bibnamefont {Zhu}}, \bibinfo
  {author} {\bibfnamefont {H.}~\bibnamefont {Wang}}, \bibinfo {author}
  {\bibfnamefont {Y.-A.}\ \bibnamefont {Chen}}, \bibinfo {author}
  {\bibfnamefont {C.-Y.}\ \bibnamefont {Lu}}, \bibinfo {author} {\bibfnamefont
  {S.}~\bibnamefont {Han}}, \ and\ \bibinfo {author} {\bibfnamefont {J.-W.}\
  \bibnamefont {Pan}},\ }\href {\doibase 10.1103/PhysRevLett.119.180511}
  {\bibfield  {journal} {\bibinfo  {journal} {Phys. Rev. Lett.}\ }\textbf
  {\bibinfo {volume} {119}},\ \bibinfo {pages} {180511} (\bibinfo {year}
  {2017})}\BibitemShut {NoStop}%
\bibitem [{\citenamefont {Quijandr\'{\i}a}\ \emph {et~al.}(2018)\citenamefont
  {Quijandr\'{\i}a}, \citenamefont {Naether}, \citenamefont {\"Ozdemir},
  \citenamefont {Nori},\ and\ \citenamefont {Zueco}}]{Quijandria2018}%
  \BibitemOpen
  \bibfield  {author} {\bibinfo {author} {\bibfnamefont {F.}~\bibnamefont
  {Quijandr\'{\i}a}}, \bibinfo {author} {\bibfnamefont {U.}~\bibnamefont
  {Naether}}, \bibinfo {author} {\bibfnamefont {S.~K.}\ \bibnamefont
  {\"Ozdemir}}, \bibinfo {author} {\bibfnamefont {F.}~\bibnamefont {Nori}}, \
  and\ \bibinfo {author} {\bibfnamefont {D.}~\bibnamefont {Zueco}},\ }\href
  {\doibase 10.1103/PhysRevA.97.053846} {\bibfield  {journal} {\bibinfo
  {journal} {Phys. Rev. A}\ }\textbf {\bibinfo {volume} {97}},\ \bibinfo
  {pages} {053846} (\bibinfo {year} {2018})}\BibitemShut {NoStop}%
\bibitem [{\citenamefont {Feng}\ \emph {et~al.}(2017)\citenamefont {Feng},
  \citenamefont {El-Ganainy},\ and\ \citenamefont {Ge}}]{feng2017non}%
  \BibitemOpen
  \bibfield  {author} {\bibinfo {author} {\bibfnamefont {L.}~\bibnamefont
  {Feng}}, \bibinfo {author} {\bibfnamefont {R.}~\bibnamefont {El-Ganainy}}, \
  and\ \bibinfo {author} {\bibfnamefont {L.}~\bibnamefont {Ge}},\ }\href
  {\doibase 10.1038/s41566-017-0031-1} {\bibfield  {journal} {\bibinfo
  {journal} {Nat. Photonics}\ }\textbf {\bibinfo {volume} {11}},\ \bibinfo
  {pages} {752} (\bibinfo {year} {2017})}\BibitemShut {NoStop}%
\bibitem [{\citenamefont {Regensburger}\ \emph {et~al.}(2012)\citenamefont
  {Regensburger}, \citenamefont {Bersch}, \citenamefont {Miri}, \citenamefont
  {Onishchukov}, \citenamefont {Christodoulides},\ and\ \citenamefont
  {Peschel}}]{regensburger2012parity}%
  \BibitemOpen
  \bibfield  {author} {\bibinfo {author} {\bibfnamefont {A.}~\bibnamefont
  {Regensburger}}, \bibinfo {author} {\bibfnamefont {C.}~\bibnamefont
  {Bersch}}, \bibinfo {author} {\bibfnamefont {M.-A.}\ \bibnamefont {Miri}},
  \bibinfo {author} {\bibfnamefont {G.}~\bibnamefont {Onishchukov}}, \bibinfo
  {author} {\bibfnamefont {D.~N.}\ \bibnamefont {Christodoulides}}, \ and\
  \bibinfo {author} {\bibfnamefont {U.}~\bibnamefont {Peschel}},\ }\href
  {\doibase 10.1038/nature11298} {\bibfield  {journal} {\bibinfo  {journal}
  {Nature}\ }\textbf {\bibinfo {volume} {488}},\ \bibinfo {pages} {167}
  (\bibinfo {year} {2012})}\BibitemShut {NoStop}%
\bibitem [{\citenamefont {Schomerus}(2013)}]{schomerus2013topologically}%
  \BibitemOpen
  \bibfield  {author} {\bibinfo {author} {\bibfnamefont {H.}~\bibnamefont
  {Schomerus}},\ }\href {\doibase 10.1364/OL.38.001912} {\bibfield  {journal}
  {\bibinfo  {journal} {Opt. Lett.}\ }\textbf {\bibinfo {volume} {38}},\
  \bibinfo {pages} {1912} (\bibinfo {year} {2013})}\BibitemShut {NoStop}%
\bibitem [{\citenamefont {Shen}\ \emph {et~al.}(2018)\citenamefont {Shen},
  \citenamefont {Zhen},\ and\ \citenamefont {Fu}}]{Shen2018}%
  \BibitemOpen
  \bibfield  {author} {\bibinfo {author} {\bibfnamefont {H.}~\bibnamefont
  {Shen}}, \bibinfo {author} {\bibfnamefont {B.}~\bibnamefont {Zhen}}, \ and\
  \bibinfo {author} {\bibfnamefont {L.}~\bibnamefont {Fu}},\ }\href {\doibase
  10.1103/PhysRevLett.120.146402} {\bibfield  {journal} {\bibinfo  {journal}
  {Phys. Rev. Lett.}\ }\textbf {\bibinfo {volume} {120}},\ \bibinfo {pages}
  {146402} (\bibinfo {year} {2018})}\BibitemShut {NoStop}%
\bibitem [{\citenamefont {Yao}\ \emph {et~al.}(2018)\citenamefont {Yao},
  \citenamefont {Song},\ and\ \citenamefont {Wang}}]{Yao2018a}%
  \BibitemOpen
  \bibfield  {author} {\bibinfo {author} {\bibfnamefont {S.}~\bibnamefont
  {Yao}}, \bibinfo {author} {\bibfnamefont {F.}~\bibnamefont {Song}}, \ and\
  \bibinfo {author} {\bibfnamefont {Z.}~\bibnamefont {Wang}},\ }\href {\doibase
  10.1103/PhysRevLett.121.136802} {\bibfield  {journal} {\bibinfo  {journal}
  {Phys. Rev. Lett.}\ }\textbf {\bibinfo {volume} {121}},\ \bibinfo {pages}
  {136802} (\bibinfo {year} {2018})}\BibitemShut {NoStop}%
\bibitem [{\citenamefont {Yao}\ and\ \citenamefont {Wang}(2018)}]{Yao2018}%
  \BibitemOpen
  \bibfield  {author} {\bibinfo {author} {\bibfnamefont {S.}~\bibnamefont
  {Yao}}\ and\ \bibinfo {author} {\bibfnamefont {Z.}~\bibnamefont {Wang}},\
  }\href {\doibase 10.1103/PhysRevLett.121.086803} {\bibfield  {journal}
  {\bibinfo  {journal} {Phys. Rev. Lett.}\ }\textbf {\bibinfo {volume} {121}},\
  \bibinfo {pages} {086803} (\bibinfo {year} {2018})}\BibitemShut {NoStop}%
\bibitem [{\citenamefont {Schomerus}\ and\ \citenamefont
  {Halpern}(2013)}]{schomerus2013parity}%
  \BibitemOpen
  \bibfield  {author} {\bibinfo {author} {\bibfnamefont {H.}~\bibnamefont
  {Schomerus}}\ and\ \bibinfo {author} {\bibfnamefont {N.~Y.}\ \bibnamefont
  {Halpern}},\ }\href {\doibase 10.1103/PhysRevLett.110.013903} {\bibfield
  {journal} {\bibinfo  {journal} {Phys. Rev. Lett.}\ }\textbf {\bibinfo
  {volume} {110}},\ \bibinfo {pages} {013903} (\bibinfo {year}
  {2013})}\BibitemShut {NoStop}%
\bibitem [{\citenamefont {Esaki}\ \emph {et~al.}(2011)\citenamefont {Esaki},
  \citenamefont {Sato}, \citenamefont {Hasebe},\ and\ \citenamefont
  {Kohmoto}}]{esaki2011edge}%
  \BibitemOpen
  \bibfield  {author} {\bibinfo {author} {\bibfnamefont {K.}~\bibnamefont
  {Esaki}}, \bibinfo {author} {\bibfnamefont {M.}~\bibnamefont {Sato}},
  \bibinfo {author} {\bibfnamefont {K.}~\bibnamefont {Hasebe}}, \ and\ \bibinfo
  {author} {\bibfnamefont {M.}~\bibnamefont {Kohmoto}},\ }\href {\doibase
  10.1103/PhysRevB.84.205128} {\bibfield  {journal} {\bibinfo  {journal} {Phys.
  Rev. B}\ }\textbf {\bibinfo {volume} {84}},\ \bibinfo {pages} {205128}
  (\bibinfo {year} {2011})}\BibitemShut {NoStop}%
\bibitem [{\citenamefont {Leykam}\ \emph {et~al.}(2017)\citenamefont {Leykam},
  \citenamefont {Bliokh}, \citenamefont {Huang}, \citenamefont {Chong},\ and\
  \citenamefont {Nori}}]{leykam2017edge}%
  \BibitemOpen
  \bibfield  {author} {\bibinfo {author} {\bibfnamefont {D.}~\bibnamefont
  {Leykam}}, \bibinfo {author} {\bibfnamefont {K.~Y.}\ \bibnamefont {Bliokh}},
  \bibinfo {author} {\bibfnamefont {C.}~\bibnamefont {Huang}}, \bibinfo
  {author} {\bibfnamefont {Y.~D.}\ \bibnamefont {Chong}}, \ and\ \bibinfo
  {author} {\bibfnamefont {F.}~\bibnamefont {Nori}},\ }\href {\doibase
  10.1103/PhysRevLett.118.040401} {\bibfield  {journal} {\bibinfo  {journal}
  {Phys. Rev. Lett.}\ }\textbf {\bibinfo {volume} {118}},\ \bibinfo {pages}
  {040401} (\bibinfo {year} {2017})}\BibitemShut {NoStop}%
\bibitem [{\citenamefont {Osterloh}\ \emph {et~al.}(2005)\citenamefont
  {Osterloh}, \citenamefont {Baig}, \citenamefont {Santos}, \citenamefont
  {Zoller},\ and\ \citenamefont {Lewenstein}}]{Osterloh2005}%
  \BibitemOpen
  \bibfield  {author} {\bibinfo {author} {\bibfnamefont {K.}~\bibnamefont
  {Osterloh}}, \bibinfo {author} {\bibfnamefont {M.}~\bibnamefont {Baig}},
  \bibinfo {author} {\bibfnamefont {L.}~\bibnamefont {Santos}}, \bibinfo
  {author} {\bibfnamefont {P.}~\bibnamefont {Zoller}}, \ and\ \bibinfo {author}
  {\bibfnamefont {M.}~\bibnamefont {Lewenstein}},\ }\href {\doibase
  10.1103/PhysRevLett.95.010403} {\bibfield  {journal} {\bibinfo  {journal}
  {Phys. Rev. Lett.}\ }\textbf {\bibinfo {volume} {95}},\ \bibinfo {pages}
  {010403} (\bibinfo {year} {2005})}\BibitemShut {NoStop}%
\bibitem [{sup()}]{supp}%
  \BibitemOpen
  \href@noop {} {}\bibinfo {note} {See Supplementary Materials for details of
  the physical realization, engineering of the dissipation, band structure,
  flying butterfly effect and dectection method.}\BibitemShut {Stop}%
\bibitem [{\citenamefont {Hofstadter}(1976)}]{hofstadter1976energy}%
  \BibitemOpen
  \bibfield  {author} {\bibinfo {author} {\bibfnamefont {D.~R.}\ \bibnamefont
  {Hofstadter}},\ }\href {\doibase 10.1103/PhysRevB.14.2239} {\bibfield
  {journal} {\bibinfo  {journal} {Phys. Rev. B}\ }\textbf {\bibinfo {volume}
  {14}},\ \bibinfo {pages} {2239} (\bibinfo {year} {1976})}\BibitemShut
  {NoStop}%
\bibitem [{\citenamefont {Azbel}(1964)}]{azbel1964energy}%
  \BibitemOpen
  \bibfield  {author} {\bibinfo {author} {\bibfnamefont {M.~Y.}\ \bibnamefont
  {Azbel}},\ }\href@noop {} {\bibfield  {journal} {\bibinfo  {journal} {Sov.
  Phys. JETP}\ }\textbf {\bibinfo {volume} {19}},\ \bibinfo {pages} {634}
  (\bibinfo {year} {1964})}\BibitemShut {NoStop}%
\bibitem [{\citenamefont {Claro}\ and\ \citenamefont
  {Wannier}(1979)}]{claro1979magnetic}%
  \BibitemOpen
  \bibfield  {author} {\bibinfo {author} {\bibfnamefont {F.~H.}\ \bibnamefont
  {Claro}}\ and\ \bibinfo {author} {\bibfnamefont {G.~H.}\ \bibnamefont
  {Wannier}},\ }\href {\doibase 10.1103/PhysRevB.19.6068} {\bibfield  {journal}
  {\bibinfo  {journal} {Phys. Rev. B}\ }\textbf {\bibinfo {volume} {19}},\
  \bibinfo {pages} {6068} (\bibinfo {year} {1979})}\BibitemShut {NoStop}%
\bibitem [{\citenamefont {Rammal}(1985)}]{rammal1985landau}%
  \BibitemOpen
  \bibfield  {author} {\bibinfo {author} {\bibfnamefont {R.}~\bibnamefont
  {Rammal}},\ }\href@noop {} {\bibfield  {journal} {\bibinfo  {journal}
  {Journal de Physique}\ }\textbf {\bibinfo {volume} {46}},\ \bibinfo {pages}
  {1345} (\bibinfo {year} {1985})}\BibitemShut {NoStop}%
\bibitem [{\citenamefont {Sheng}\ \emph {et~al.}(2006)\citenamefont {Sheng},
  \citenamefont {Weng}, \citenamefont {Sheng},\ and\ \citenamefont
  {Haldane}}]{sheng2006quantum}%
  \BibitemOpen
  \bibfield  {author} {\bibinfo {author} {\bibfnamefont {D.~N.}\ \bibnamefont
  {Sheng}}, \bibinfo {author} {\bibfnamefont {Z.~Y.}\ \bibnamefont {Weng}},
  \bibinfo {author} {\bibfnamefont {L.}~\bibnamefont {Sheng}}, \ and\ \bibinfo
  {author} {\bibfnamefont {F.~D.~M.}\ \bibnamefont {Haldane}},\ }\href
  {\doibase 10.1103/PhysRevLett.97.036808} {\bibfield  {journal} {\bibinfo
  {journal} {Phys. Rev. Lett.}\ }\textbf {\bibinfo {volume} {97}},\ \bibinfo
  {pages} {036808} (\bibinfo {year} {2006})}\BibitemShut {NoStop}%
\bibitem [{\citenamefont {Prodan}(2009)}]{prodan2009robustness}%
  \BibitemOpen
  \bibfield  {author} {\bibinfo {author} {\bibfnamefont {E.}~\bibnamefont
  {Prodan}},\ }\href {\doibase 10.1103/PhysRevB.80.125327} {\bibfield
  {journal} {\bibinfo  {journal} {Phys. Rev. B}\ }\textbf {\bibinfo {volume}
  {80}},\ \bibinfo {pages} {125327} (\bibinfo {year} {2009})}\BibitemShut
  {NoStop}%
\bibitem [{\citenamefont {Wang}\ \emph {et~al.}(2016)\citenamefont {Wang},
  \citenamefont {Yang}, \citenamefont {Hu}, \citenamefont {Xue},\ and\
  \citenamefont {Wu}}]{Wang2016}%
  \BibitemOpen
  \bibfield  {author} {\bibinfo {author} {\bibfnamefont {Y.-P.}\ \bibnamefont
  {Wang}}, \bibinfo {author} {\bibfnamefont {W.-L.}\ \bibnamefont {Yang}},
  \bibinfo {author} {\bibfnamefont {Y.}~\bibnamefont {Hu}}, \bibinfo {author}
  {\bibfnamefont {Z.-Y.}\ \bibnamefont {Xue}}, \ and\ \bibinfo {author}
  {\bibfnamefont {Y.}~\bibnamefont {Wu}},\ }\href {\doibase
  10.1038/npjqi.2016.15} {\bibfield  {journal} {\bibinfo  {journal} {npj
  Quantum Inf.}\ }\textbf {\bibinfo {volume} {2}},\ \bibinfo {pages} {16015}
  (\bibinfo {year} {2016})}\BibitemShut {NoStop}%
\bibitem [{\citenamefont {Fukui}\ \emph {et~al.}(2008)\citenamefont {Fukui},
  \citenamefont {Fujiwara},\ and\ \citenamefont
  {Hatsugai}}]{fukui2008topological}%
  \BibitemOpen
  \bibfield  {author} {\bibinfo {author} {\bibfnamefont {T.}~\bibnamefont
  {Fukui}}, \bibinfo {author} {\bibfnamefont {T.}~\bibnamefont {Fujiwara}}, \
  and\ \bibinfo {author} {\bibfnamefont {Y.}~\bibnamefont {Hatsugai}},\
  }\href@noop {} {\bibfield  {journal} {\bibinfo  {journal} {Journal of the
  Physical Society of Japan}\ }\textbf {\bibinfo {volume} {77}},\ \bibinfo
  {pages} {123705} (\bibinfo {year} {2008})}\BibitemShut {NoStop}%
\bibitem [{\citenamefont {Lee}(2016)}]{lee2016anomalous}%
  \BibitemOpen
  \bibfield  {author} {\bibinfo {author} {\bibfnamefont {T.~E.}\ \bibnamefont
  {Lee}},\ }\href {\doibase 10.1103/PhysRevLett.116.133903} {\bibfield
  {journal} {\bibinfo  {journal} {Phys. Rev. Lett.}\ }\textbf {\bibinfo
  {volume} {116}},\ \bibinfo {pages} {133903} (\bibinfo {year}
  {2016})}\BibitemShut {NoStop}%
\bibitem [{\citenamefont {Zeuner}\ \emph {et~al.}(2015)\citenamefont {Zeuner},
  \citenamefont {Rechtsman}, \citenamefont {Plotnik}, \citenamefont {Lumer},
  \citenamefont {Nolte}, \citenamefont {Rudner}, \citenamefont {Segev},\ and\
  \citenamefont {Szameit}}]{zeuner2015observation}%
  \BibitemOpen
  \bibfield  {author} {\bibinfo {author} {\bibfnamefont {J.~M.}\ \bibnamefont
  {Zeuner}}, \bibinfo {author} {\bibfnamefont {M.~C.}\ \bibnamefont
  {Rechtsman}}, \bibinfo {author} {\bibfnamefont {Y.}~\bibnamefont {Plotnik}},
  \bibinfo {author} {\bibfnamefont {Y.}~\bibnamefont {Lumer}}, \bibinfo
  {author} {\bibfnamefont {S.}~\bibnamefont {Nolte}}, \bibinfo {author}
  {\bibfnamefont {M.~S.}\ \bibnamefont {Rudner}}, \bibinfo {author}
  {\bibfnamefont {M.}~\bibnamefont {Segev}}, \ and\ \bibinfo {author}
  {\bibfnamefont {A.}~\bibnamefont {Szameit}},\ }\href {\doibase
  10.1103/PhysRevLett.115.040402} {\bibfield  {journal} {\bibinfo  {journal}
  {Phys. Rev. Lett.}\ }\textbf {\bibinfo {volume} {115}},\ \bibinfo {pages}
  {040402} (\bibinfo {year} {2015})}\BibitemShut {NoStop}%
\bibitem [{\citenamefont {Malzard}\ \emph {et~al.}(2015)\citenamefont
  {Malzard}, \citenamefont {Poli},\ and\ \citenamefont
  {Schomerus}}]{malzard2015topologically}%
  \BibitemOpen
  \bibfield  {author} {\bibinfo {author} {\bibfnamefont {S.}~\bibnamefont
  {Malzard}}, \bibinfo {author} {\bibfnamefont {C.}~\bibnamefont {Poli}}, \
  and\ \bibinfo {author} {\bibfnamefont {H.}~\bibnamefont {Schomerus}},\ }\href
  {\doibase 10.1103/PhysRevLett.115.200402} {\bibfield  {journal} {\bibinfo
  {journal} {Phys. Rev. Lett.}\ }\textbf {\bibinfo {volume} {115}},\ \bibinfo
  {pages} {200402} (\bibinfo {year} {2015})}\BibitemShut {NoStop}%
\bibitem [{cai()}]{caibutterflyeffect}%
  \BibitemOpen
  \href@noop {} {}\bibinfo {note} {Jia-Qi Cai, Zheng-Yuan Xue, Ming Gong,
  Guang-Can Guo, and Yong Hu, Hofstadter butterfly in non-Hermitian models. In
  preparation.}\BibitemShut {Stop}%
\end{thebibliography}

\begin{thebibliography}{8}%
	\makeatletter
	\providecommand \@ifxundefined [1]{%
		\@ifx{#1\undefined}
	}%
	\providecommand \@ifnum [1]{%
		\ifnum #1\expandafter \@firstoftwo
		\else \expandafter \@secondoftwo
		\fi
	}%
	\providecommand \@ifx [1]{%
		\ifx #1\expandafter \@firstoftwo
		\else \expandafter \@secondoftwo
		\fi
	}%
	\providecommand \natexlab [1]{#1}%
	\providecommand \enquote  [1]{``#1''}%
	\providecommand \bibnamefont  [1]{#1}%
	\providecommand \bibfnamefont [1]{#1}%
	\providecommand \citenamefont [1]{#1}%
	\providecommand \href@noop [0]{\@secondoftwo}%
	\providecommand \href [0]{\begingroup \@sanitize@url \@href}%
	\providecommand \@href[1]{\@@startlink{#1}\@@href}%
	\providecommand \@@href[1]{\endgroup#1\@@endlink}%
	\providecommand \@sanitize@url [0]{\catcode `\\12\catcode `\$12\catcode
		`\&12\catcode `\#12\catcode `\^12\catcode `\_12\catcode `\%12\relax}%
	\providecommand \@@startlink[1]{}%
	\providecommand \@@endlink[0]{}%
	\providecommand \url  [0]{\begingroup\@sanitize@url \@url }%
	\providecommand \@url [1]{\endgroup\@href {#1}{\urlprefix }}%
	\providecommand \urlprefix  [0]{URL }%
	\providecommand \Eprint [0]{\href }%
	\providecommand \doibase [0]{http://dx.doi.org/}%
	\providecommand \selectlanguage [0]{\@gobble}%
	\providecommand \bibinfo  [0]{\@secondoftwo}%
	\providecommand \bibfield  [0]{\@secondoftwo}%
	\providecommand \translation [1]{[#1]}%
	\providecommand \BibitemOpen [0]{}%
	\providecommand \bibitemStop [0]{}%
	\providecommand \bibitemNoStop [0]{.\EOS\space}%
	\providecommand \EOS [0]{\spacefactor3000\relax}%
	\providecommand \BibitemShut  [1]{\csname bibitem#1\endcsname}%
	\let\auto@bib@innerbib\@empty
	%</preamble>
	\bibitem [{\citenamefont {Paik}\ \emph {et~al.}(2011)\citenamefont {Paik},
		\citenamefont {Schuster}, \citenamefont {Bishop}, \citenamefont {Kirchmair},
		\citenamefont {Catelani}, \citenamefont {Sears}, \citenamefont {Johnson},
		\citenamefont {Reagor}, \citenamefont {Frunzio}, \citenamefont {Glazman},
		\citenamefont {Girvin}, \citenamefont {Devoret},\ and\ \citenamefont
		{Schoelkopf}}]{Paik2011}%
	\BibitemOpen
	\bibfield  {author} {\bibinfo {author} {\bibfnamefont {H.}~\bibnamefont
			{Paik}}, \bibinfo {author} {\bibfnamefont {D.~I.}\ \bibnamefont {Schuster}},
		\bibinfo {author} {\bibfnamefont {L.~S.}\ \bibnamefont {Bishop}}, \bibinfo
		{author} {\bibfnamefont {G.}~\bibnamefont {Kirchmair}}, \bibinfo {author}
		{\bibfnamefont {G.}~\bibnamefont {Catelani}}, \bibinfo {author}
		{\bibfnamefont {A.~P.}\ \bibnamefont {Sears}}, \bibinfo {author}
		{\bibfnamefont {B.~R.}\ \bibnamefont {Johnson}}, \bibinfo {author}
		{\bibfnamefont {M.~J.}\ \bibnamefont {Reagor}}, \bibinfo {author}
		{\bibfnamefont {L.}~\bibnamefont {Frunzio}}, \bibinfo {author} {\bibfnamefont
			{L.~I.}\ \bibnamefont {Glazman}}, \bibinfo {author} {\bibfnamefont {S.~M.}\
			\bibnamefont {Girvin}}, \bibinfo {author} {\bibfnamefont {M.~H.}\
			\bibnamefont {Devoret}}, \ and\ \bibinfo {author} {\bibfnamefont {R.~J.}\
			\bibnamefont {Schoelkopf}},\ }\href {\doibase 10.1103/PhysRevLett.107.240501}
	{\bibfield  {journal} {\bibinfo  {journal} {Phys. Rev. Lett.}\ }\textbf
		{\bibinfo {volume} {107}},\ \bibinfo {pages} {240501} (\bibinfo {year}
		{2011})}\BibitemShut {NoStop}%
	\bibitem [{\citenamefont {Reagor}\ \emph {et~al.}(2013)\citenamefont {Reagor},
		\citenamefont {Paik}, \citenamefont {Catelani}, \citenamefont {Sun},
		\citenamefont {Axline}, \citenamefont {Holland}, \citenamefont {Pop},
		\citenamefont {Masluk}, \citenamefont {Brecht}, \citenamefont {Frunzio},
		\citenamefont {Devoret}, \citenamefont {Glazman},\ and\ \citenamefont
		{Schoelkopf}}]{reagor2013reaching}%
	\BibitemOpen
	\bibfield  {author} {\bibinfo {author} {\bibfnamefont {M.}~\bibnamefont
			{Reagor}}, \bibinfo {author} {\bibfnamefont {H.}~\bibnamefont {Paik}},
		\bibinfo {author} {\bibfnamefont {G.}~\bibnamefont {Catelani}}, \bibinfo
		{author} {\bibfnamefont {L.}~\bibnamefont {Sun}}, \bibinfo {author}
		{\bibfnamefont {C.}~\bibnamefont {Axline}}, \bibinfo {author} {\bibfnamefont
			{E.}~\bibnamefont {Holland}}, \bibinfo {author} {\bibfnamefont {I.~M.}\
			\bibnamefont {Pop}}, \bibinfo {author} {\bibfnamefont {N.~A.}\ \bibnamefont
			{Masluk}}, \bibinfo {author} {\bibfnamefont {T.}~\bibnamefont {Brecht}},
		\bibinfo {author} {\bibfnamefont {L.}~\bibnamefont {Frunzio}}, \bibinfo
		{author} {\bibfnamefont {M.~H.}\ \bibnamefont {Devoret}}, \bibinfo {author}
		{\bibfnamefont {L.}~\bibnamefont {Glazman}}, \ and\ \bibinfo {author}
		{\bibfnamefont {R.~J.}\ \bibnamefont {Schoelkopf}},\ }\href {\doibase
		10.1063/1.4807015} {\bibfield  {journal} {\bibinfo  {journal} {Appl. Phys.
				Lett}\ }\textbf {\bibinfo {volume} {102}},\ \bibinfo {pages} {192604}
		(\bibinfo {year} {2013})}\BibitemShut {NoStop}%
	\bibitem [{\citenamefont {Sirois}\ \emph {et~al.}(2015)\citenamefont {Sirois},
		\citenamefont {Castellanos-Beltran}, \citenamefont {DeFeo}, \citenamefont
		{Ranzani}, \citenamefont {Lecocq}, \citenamefont {Simmonds}, \citenamefont
		{Teufel},\ and\ \citenamefont {Aumentado}}]{Sirois2015}%
	\BibitemOpen
	\bibfield  {author} {\bibinfo {author} {\bibfnamefont {A.~J.}\ \bibnamefont
			{Sirois}}, \bibinfo {author} {\bibfnamefont {M.~A.}\ \bibnamefont
			{Castellanos-Beltran}}, \bibinfo {author} {\bibfnamefont {M.~P.}\
			\bibnamefont {DeFeo}}, \bibinfo {author} {\bibfnamefont {L.}~\bibnamefont
			{Ranzani}}, \bibinfo {author} {\bibfnamefont {F.}~\bibnamefont {Lecocq}},
		\bibinfo {author} {\bibfnamefont {R.~W.}\ \bibnamefont {Simmonds}}, \bibinfo
		{author} {\bibfnamefont {J.~D.}\ \bibnamefont {Teufel}}, \ and\ \bibinfo
		{author} {\bibfnamefont {J.}~\bibnamefont {Aumentado}},\ }\href {\doibase
		10.1063/1.4919759} {\bibfield  {journal} {\bibinfo  {journal} {Appl. Phys
				Lett.}\ }\textbf {\bibinfo {volume} {106}},\ \bibinfo {pages} {172603}
		(\bibinfo {year} {2015})}\BibitemShut {NoStop}%
	\bibitem [{\citenamefont {Reagor}\ \emph {et~al.}(2016)\citenamefont {Reagor},
		\citenamefont {Pfaff}, \citenamefont {Axline}, \citenamefont {Heeres},
		\citenamefont {Ofek}, \citenamefont {Sliwa}, \citenamefont {Holland},
		\citenamefont {Wang}, \citenamefont {Blumoff}, \citenamefont {Chou},
		\citenamefont {Hatridge}, \citenamefont {Frunzio}, \citenamefont {Devoret},
		\citenamefont {Jiang},\ and\ \citenamefont {Schoelkopf}}]{reagor2016quantum}%
	\BibitemOpen
	\bibfield  {author} {\bibinfo {author} {\bibfnamefont {M.}~\bibnamefont
			{Reagor}}, \bibinfo {author} {\bibfnamefont {W.}~\bibnamefont {Pfaff}},
		\bibinfo {author} {\bibfnamefont {C.}~\bibnamefont {Axline}}, \bibinfo
		{author} {\bibfnamefont {R.~W.}\ \bibnamefont {Heeres}}, \bibinfo {author}
		{\bibfnamefont {N.}~\bibnamefont {Ofek}}, \bibinfo {author} {\bibfnamefont
			{K.}~\bibnamefont {Sliwa}}, \bibinfo {author} {\bibfnamefont
			{E.}~\bibnamefont {Holland}}, \bibinfo {author} {\bibfnamefont
			{C.}~\bibnamefont {Wang}}, \bibinfo {author} {\bibfnamefont {J.}~\bibnamefont
			{Blumoff}}, \bibinfo {author} {\bibfnamefont {K.}~\bibnamefont {Chou}},
		\bibinfo {author} {\bibfnamefont {M.~J.}\ \bibnamefont {Hatridge}}, \bibinfo
		{author} {\bibfnamefont {L.}~\bibnamefont {Frunzio}}, \bibinfo {author}
		{\bibfnamefont {M.~H.}\ \bibnamefont {Devoret}}, \bibinfo {author}
		{\bibfnamefont {L.}~\bibnamefont {Jiang}}, \ and\ \bibinfo {author}
		{\bibfnamefont {R.~J.}\ \bibnamefont {Schoelkopf}},\ }\href {\doibase
		10.1103/PhysRevB.94.014506} {\bibfield  {journal} {\bibinfo  {journal} {Phys.
				Rev. B}\ }\textbf {\bibinfo {volume} {94}},\ \bibinfo {pages} {014506}
		(\bibinfo {year} {2016})}\BibitemShut {NoStop}%
	\bibitem [{\citenamefont {Owens}\ \emph {et~al.}(2018)\citenamefont {Owens},
		\citenamefont {LaChapelle}, \citenamefont {Saxberg}, \citenamefont
		{Anderson}, \citenamefont {Ma}, \citenamefont {Simon},\ and\ \citenamefont
		{Schuster}}]{owens2018quarter}%
	\BibitemOpen
	\bibfield  {author} {\bibinfo {author} {\bibfnamefont {C.}~\bibnamefont
			{Owens}}, \bibinfo {author} {\bibfnamefont {A.}~\bibnamefont {LaChapelle}},
		\bibinfo {author} {\bibfnamefont {B.}~\bibnamefont {Saxberg}}, \bibinfo
		{author} {\bibfnamefont {B.~M.}\ \bibnamefont {Anderson}}, \bibinfo {author}
		{\bibfnamefont {R.}~\bibnamefont {Ma}}, \bibinfo {author} {\bibfnamefont
			{J.}~\bibnamefont {Simon}}, \ and\ \bibinfo {author} {\bibfnamefont {D.~I.}\
			\bibnamefont {Schuster}},\ }\href {\doibase 10.1103/PhysRevA.97.013818}
	{\bibfield  {journal} {\bibinfo  {journal} {Phys. Rev. A}\ }\textbf {\bibinfo
			{volume} {97}},\ \bibinfo {pages} {013818} (\bibinfo {year}
		{2018})}\BibitemShut {NoStop}%
	\bibitem [{\citenamefont {Lu}\ \emph {et~al.}(2018)\citenamefont {Lu} \emph
		{et~al.}}]{lu2018spontaneous}%
	\BibitemOpen
	\bibfield  {author} {\bibinfo {author} {\bibfnamefont {Y.-K.}\ \bibnamefont
			{Lu}} \emph {et~al.},\ }\href {\doibase
		https://doi.org/10.1016/j.scib.2018.07.020} {\bibfield  {journal} {\bibinfo
			{journal} {Sci. Bull.}\ }\textbf {\bibinfo {volume} {63}},\ \bibinfo {pages}
		{1096} (\bibinfo {year} {2018})}\BibitemShut {NoStop}%
	\bibitem [{\citenamefont {Quijandr\'{\i}a}\ \emph {et~al.}(2018)\citenamefont
		{Quijandr\'{\i}a}, \citenamefont {Naether}, \citenamefont {\"Ozdemir},
		\citenamefont {Nori},\ and\ \citenamefont {Zueco}}]{Quijandria2018}%
	\BibitemOpen
	\bibfield  {author} {\bibinfo {author} {\bibfnamefont {F.}~\bibnamefont
			{Quijandr\'{\i}a}}, \bibinfo {author} {\bibfnamefont {U.}~\bibnamefont
			{Naether}}, \bibinfo {author} {\bibfnamefont {S.~K.}\ \bibnamefont
			{\"Ozdemir}}, \bibinfo {author} {\bibfnamefont {F.}~\bibnamefont {Nori}}, \
		and\ \bibinfo {author} {\bibfnamefont {D.}~\bibnamefont {Zueco}},\ }\href
	{\doibase 10.1103/PhysRevA.97.053846} {\bibfield  {journal} {\bibinfo
			{journal} {Phys. Rev. A}\ }\textbf {\bibinfo {volume} {97}},\ \bibinfo
		{pages} {053846} (\bibinfo {year} {2018})}\BibitemShut {NoStop}%
	\bibitem [{\citenamefont {Wen}\ and\ \citenamefont
		{Zee}(1989)}]{wen1989winding}%
	\BibitemOpen
	\bibfield  {author} {\bibinfo {author} {\bibfnamefont {X.}~\bibnamefont
			{Wen}}\ and\ \bibinfo {author} {\bibfnamefont {A.}~\bibnamefont {Zee}},\
	}\href@noop {} {\bibfield  {journal} {\bibinfo  {journal} {Nuclear Physics
				B}\ }\textbf {\bibinfo {volume} {316}},\ \bibinfo {pages} {641} (\bibinfo
		{year} {1989})}\BibitemShut {NoStop}%
\end{thebibliography}


\end{document}